\documentclass[
 reprint,amsmath,amssymb,aps,superscriptaddress,
]{revtex4-2}
\usepackage{graphicx} 
\setcitestyle{super}
\usepackage{physics}
\usepackage{float}
\usepackage{multibib}
\usepackage{lineno}
\usepackage{xcolor}
\usepackage{braket}
\usepackage{textcomp}

\usepackage{amssymb}
\usepackage{mathtools}

\usepackage{times}
\usepackage{float}

\newcites{SI}{Supplementary}

\begin{document}

\title{
A two-dimensional piezo-optomechanical transducer
}

\author{Tian Xie}
    \thanks{These authors contributed equally to this work.}
    \affiliation{Department of Applied Physics and E.L. Ginzton Laboratory, Stanford University, Stanford, CA, USA}
\author{Nelson Ooi}
    \thanks{These authors contributed equally to this work.}    
    \affiliation{Department of Electrical Engineering and E.L. Ginzton Laboratory, Stanford University, Stanford, CA, USA}
\author{Linus Woodard}   
    \affiliation{Department of Applied Physics and E.L. Ginzton Laboratory, Stanford University, Stanford, CA, USA}
\author{Sultan Malik}   
    \affiliation{Department of Applied Physics and E.L. Ginzton Laboratory, Stanford University, Stanford, CA, USA}
\author{Felix Mayor}   
    \affiliation{Department of Applied Physics and E.L. Ginzton Laboratory, Stanford University, Stanford, CA, USA}
\author{Oliver A. Hitchcock}   
    \affiliation{Department of Physics, Stanford University, Stanford, California 94305, USA} 
\author{André G. Primo}   
    \affiliation{Hybrid Photonics Laboratory, École Polytechnique Fédérale de Lausanne (EPFL),  CH-1015, Switzerland}   
\author{Wentao Jiang}   
    \affiliation{Department of Applied Physics and E.L. Ginzton Laboratory, Stanford University, Stanford, CA, USA}
\author{Samuel Gyger}   
    \affiliation{Department of Applied Physics and E.L. Ginzton Laboratory, Stanford University, Stanford, CA, USA}
\author{Amir H. Safavi-Naeini}
    \email[]{safavi@stanford.edu}
    \affiliation{Department of Applied Physics and E.L. Ginzton Laboratory, Stanford University, Stanford, CA, USA}
    
\begin{abstract}

Optical quantum networks provide a natural route for connecting distant superconducting quantum processors, enabling distributed quantum computation, sensing, and communication. Piezo-optomechanical transducers are among the leading candidates for scalable microwave-to-optical quantum interfaces. However, prior one-dimensional piezo-optomechanical transducers remain limited by optical-absorption-induced heating and the resulting thermal noise. Two-dimensional optomechanical crystals offer substantially improved thermalization from better thermal anchoring, but their structural complexity has so far hindered the realization of a fully integrated two-dimensional transducer. Here, we overcome this challenge with a new design strategy based on band structure engineering. We fabricate the devices and experimentally characterize the response, measuring an electromechanical damping rate of 4.6 kHz at room temperature and electromechanical coupling rate of 0.17 MHz by wire-bonding to a multi-mode microwave resonator at 10 mK. Bidirectional transduction is performed with a calibrated internal efficiency of 0.85\% in the continuous-wave operation, alongside pulsed photon-phonon pair generation near its quantum ground state. Our results represent a significant step toward high-efficiency and low-noise transducers for entangling remote superconducting qubits.

\end{abstract}

\maketitle

\section*{Introduction}
Superconducting quantum circuits are among the most promising platforms for scalable quantum computation \cite{arute2019quantum,Sivak2023}. Since superconducting circuits operate at microwave frequencies and millikelvin temperatures, direct long-distance transmission of microwave quantum states is challenging due to thermal noise and photon loss at room temperature. Microwave-to-optical (M2O) quantum transducers provide a route to overcome this limitation by coherently converting microwave quantum states into optical photons, which can propagate through optical fiber over long distances with low loss and negligible thermal mode occupation at room temperature \cite{han2021,lambert2020}. Toward this goal, several physical platforms have been explored, including electro-optic devices \cite{Warner2025,Charles2025,Sahu2023,Shen2024}, optomechanical systems \cite{Jiang2023,Delaney2022,Meesala2024,Zhao2025}, and atomic or solid-state spin ensembles \cite{Kumar2023,Xie2025}. Among different approaches, piezo-optomechanical transducers have shown particular promise\cite{sekine2026microwavetoopticalquantumtransductionphotons}; here, mechanical phonons mediate the interconversion between microwave photons and optical photons via the piezoelectric and optomechanical effects\cite{Eichenfield2009,Jiang2023}. Recent demonstrations have shown quantum-enabled microwave-to-optical conversion \cite{Zhao2025}, microwave–optical photon-pair generation \cite{Jiang2023,Meesala2024}, microwave–optical entanglement \cite{Srujan2024}, and optical readout of superconducting qubits \cite{Mirhosseini2020}.

While existing piezo-optomechanical transducers have shown promising proof-of-principle demonstrations, their performance remains limited by optical absorption-induced heating within suspended one-dimensional nanobeam-type optomechanical crystals (1D OMC) due to poor thermal anchoring\cite{Jiang2023,Zhao2025,Srujan2024}. This constrains the optical pump power and pulse repetition rates for practical networking applications \cite{weaver2025scalable,Wang2022,Krastanov2021}. Recent studies show progress on release-free OMCs where heat could damp into a bulk substrate \cite{Burger2025,kolvik2025optomechanical}, but a demonstration of a high-efficiency integrated transducer with coupling rates approaching prior benchmarks remains to be explored \cite{burger2026}. 

An alternative approach, which we pursue here, retains the released geometry but replaces the one-dimensional nanobeam with a two-dimensional optomechanical crystal (2D OMC) cavity \cite{Amir2010,Ren2020}. In a 2D OMC, the cavity is embedded in a membrane anchored to the substrate along its full perimeter: a patterned phononic shield confines the coherent gigahertz phonons, while absorption-induced heat escapes through the surrounding slab. This bandgap-engineered thermal anchoring allows 2D OMCs to combine strong optomechanical coupling and high optical quality factors with far better thermalization, supporting higher repetition rates and lower added noise than their 1D counterparts \cite{Ren2020,Sonar2025,mayor2025high,Chen2026}. Two obstacles have nonetheless kept 2D OMCs out of integrated transducers. The first is the mechanical frequency. 2D OMC demonstrations historically operated at or above 10~GHz \cite{Ren2020,Sonar2025}, where the phonon density of states, growing as the cube of frequency, fills the spectrum with spurious mechanical modes, and where a matched piezoelectric element requires electrode dimensions at the edge of fabrication limits. Meanwhile, the optical pump power needed for a given optomechanical cooperativity also grows with mechanical frequency. The recently developed b-dagger crystal \cite{mayor2025high} alleviated these constraints by lowering the breathing-mode frequency to $\sim$7~GHz. 

\begin{figure*}
\centering
\includegraphics[width=0.8\linewidth]{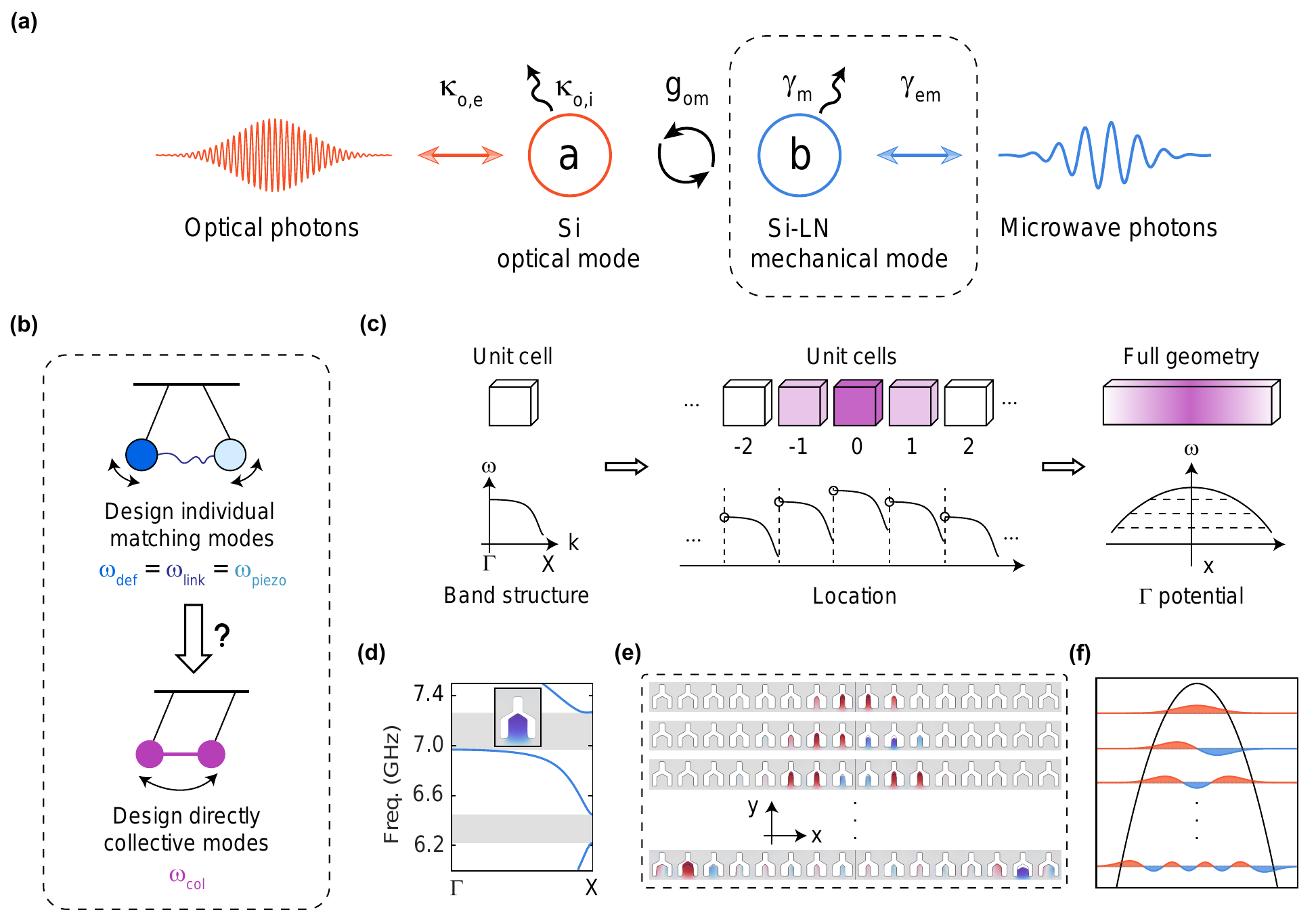}
\caption{\label{Fig1}
\textbf{Design strategy based on a mechanical $\Gamma$-potential.}
\textbf{a}, Schematic of the microwave-to-optical transduction in a piezo-optomechanical system. Microwave photons couple to a hybridized silicon-lithium niobate (Si-LN) mechanical mode, labeled mode (b), with an electromechanical damping rate $\gamma_{\mathrm{em}}$. The intrinsic mechanical decay rate is denoted by $\gamma_m$. The Si-LN mode interacts with a co-localized optical cavity mode in silicon, labeled mode (a),  with an optomechanical coupling rate at $g_\mathrm{om}$. The optical mode has intrinsic and external decay rates $\kappa_{o,i}$ and $\kappa_{o,e}$, respectively. 
\textbf{b}, Zoomed-in illustration of the hybridized Si-LN mode. In previous approaches, the defect, link, and LN mechanical modes are designed separately and then frequency-matched to enable hybridization. Here, we introduce a design strategy that directly engineers the collective mechanical mode.
\textbf{c}, Illustration of the $\Gamma$-potential design strategy. Starting from a single unit cell, the mechanical band can be shifted upward or downward by modifying the unit cell geometry. Connecting the local $\Gamma$-point frequencies across the device forms an effective $\Gamma$-potential, where each horizontal dashed line represents a supported mechanical eigenmode. 
\textbf{d}, Optomechanical unit cell used in this work (inset) and its mechanical breathing mode band selected for $\Gamma$-potential engineering. Gray regions show the band gaps separating the band of interest from adjacent modes.
\textbf{e}, Simulated mechanical eigenmodes formed by the engineered $\Gamma$-potential. The plotted color shows the displacement component along the y direction. Only the central row is plotted, omitting the 2D phononic-photonic shields. The structure is reflection symmetric about the x-axis. 
\textbf{f}, Bound-state solutions of a quantum harmonic oscillator in a parabolic potential, plotted with an inverted vertical axis. This is shown as an analogy to the discrete collective modes supported by the $\Gamma$-potential.
}
\end{figure*}

The second obstacle is the electromechanical interface itself. Phonons must be routed between the OMC cavity and a compact, electrically actuated mechanical element without degrading the optical mode. In 1D transducers, this is accomplished with a frequency-matched phonon waveguide linking the defect to the piezoelectric resonator \cite{Mirhosseini2020,Jiang2023,Chiappina2023}, but this strategy transfers poorly to two dimensions. The optomechanically favorable breathing mode has a very flat band extending past the high-symmetry $\Gamma$ point, so any connecting structure must not perturb the geometry too quickly to preserve the optical bandgap and sustain a high optical quality factor. Also, the 2D geometry admits far more mechanical modes and propagation directions through which phonons leak. Matching several discrete modes is already acutely sensitive to fabrication disorder in 1D \cite{patel2017}, and this sensitivity only compounds with the added complexity and flatness of the phononic bands. Robust electromechanical coupling in a 2D OMC has therefore remained an outstanding challenge, and has not been realized in any 2D OMC to date.

\begin{figure*}
\centering
\includegraphics[width=1\linewidth]{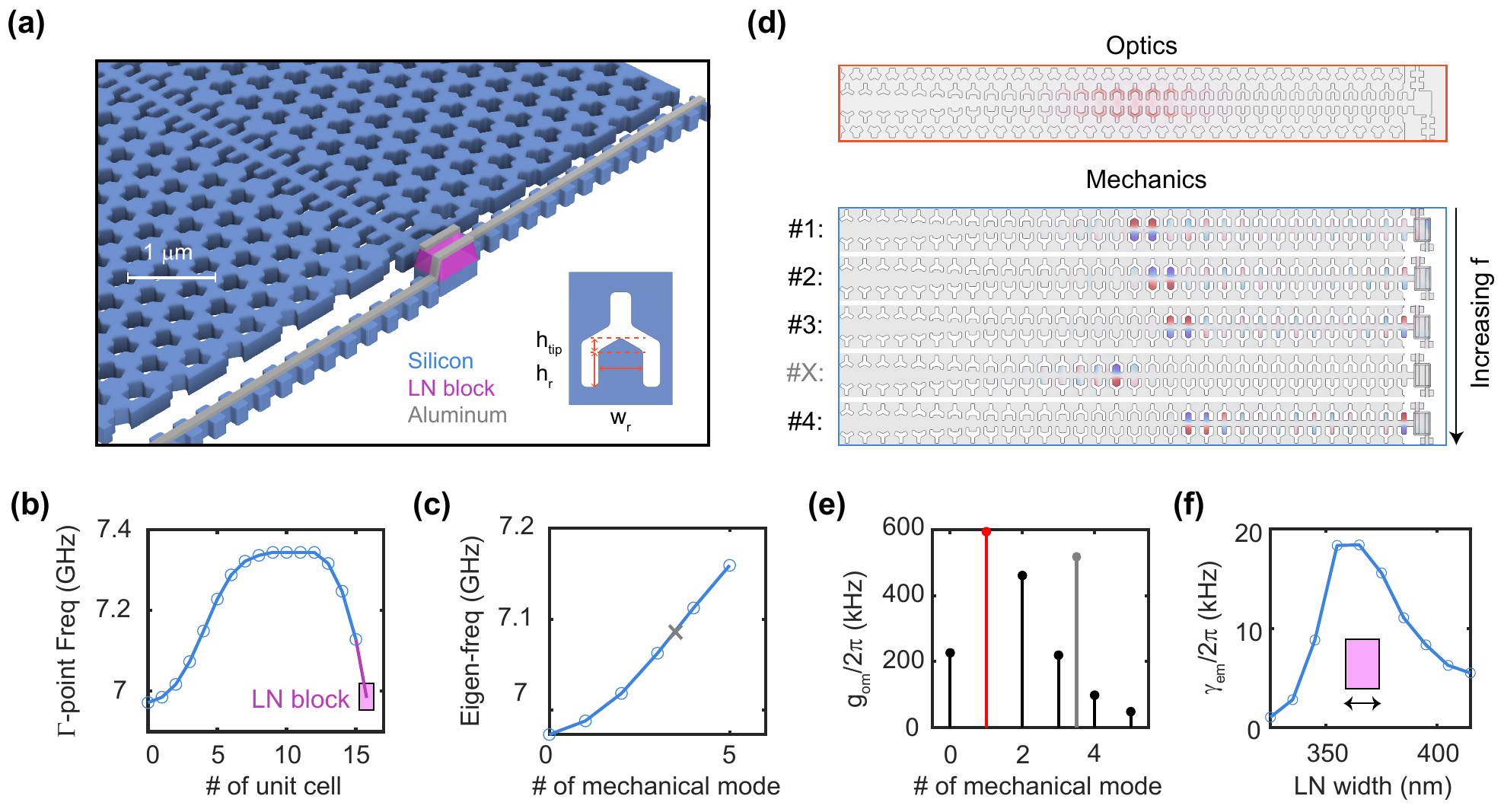}
\caption{\label{Fig2}
\textbf{2D piezo-optomechanical transducer design based on a $\Gamma$-potential.}
\textbf{a}, Model of a 2D piezo-optomechanical transducer with integrated LN and electrodes. The zoomed-in inset shows the nominal unit cell and the key geometric parameters used to tune the mechanical band structure. 
\textbf{b}, Engineered $\Gamma$-potential used in the full transducer design with the last unit cell replaced by a LN block. The potential is designed to be asymmetric to increase tolerance against fabrication disorders, as discussed in the main text.
\textbf{c}, Simulated eigenfrequencies of the collective mechanical modes supported by the $\Gamma$-potential.
\textbf{d}, Simulated optical and mechanical mode profiles of the full transducer. The mechanical modes show progressively reduced spatial extent for increasing mode index, consistent with the $\Gamma$-potential picture. 
\textbf{e}, Simulated single-photon optomechanical coupling rates for the collective mechanical modes.
\textbf{f}, Simulated electromechanical damping rate at different width of the x-cut LN block for mode 1 at a fixed LN length of 540 nm. The calculation assumes a microwave probe with 50 $\Omega$ impedance. See Supplementary Materials for details. 
}
\end{figure*}

In this work, we introduce a design strategy that overcomes this challenge by engineering the mechanical band structure of the 2D OMC itself. Spatially grading the $\Gamma$-point frequency of the breathing-mode band across the crystal creates an effective potential -- a ``$\Gamma$-potential'' -- whose bound collective mechanical modes extend from the optical cavity to a piezoelectric element while retaining large optomechanical coupling.  Because these modes are a property of the structure as a whole rather than of several frequency-matched components, the resulting interface is inherently tolerant to fabrication disorder. We validate the strategy in three stages. First, we used finite-element simulations of fully-integrated piezo-optomechanical transducers to optimize spatially-extended mechanical modes with large optomechanical and electromechanical coupling. Second, room-temperature measurements of 2D OMC cavity-only devices confirm the expected multi-mode mechanical spectrum, large optomechanical coupling and high optical quality factors. Third, integrating a transfer-printed lithium niobate block, we demonstrate microwave-to-optical conversion at room temperature and, at millikelvin temperatures, presenting bidirectional transduction with an internal efficiency of 0.85\% in continuous wave operation, alongside pulsed photon-phonon pair generation near the mechanical ground state. These results establish 2D OMCs as a practical platform for thermally robust, high-efficiency, and low-noise quantum transducers linking remote superconducting circuits. 

\section*{Design Strategy}

To motivate our design strategy, consider the electromechanical interface of a piezo-optomechanical transducer (see Fig. \ref{Fig1}a): incoming microwave photons must couple to a mechanical mode, with an electromechanical damping rate $\gamma_{\mathrm{em}}$, that simultaneously maintains a large optomechanical coupling $g_{\mathrm{om}}$ to the optical mode of the OMC cavity. Previous transducers realize such a mode by hybridizing discrete mechanical components, typically a localized defect mode at $\omega_{\mathrm{def}}$, an intermediate link mode at $\omega_{\mathrm{link}}$, and a piezoelectric mode at $\omega_{\mathrm{piezo}}$, which must be frequency-matched ($\omega_{\mathrm{def}} = \omega_{\mathrm{link}} = \omega_{\mathrm{piezo}}$) to within their hybridization bandwidths while retaining sufficient mode-profile overlap (Fig. \ref{Fig1}b) \cite{Chiappina2023,Jiang2023}. It is this group of simultaneous frequency matching conditions that makes the interface fragile to fabrication disorder. Here we take the opposite approach: rather than assembling separately designed modes (with very different characters) and relying on their hybridization, we directly engineer the collective mechanical mode of the structure as a whole.

In 2D OMCs, an in-plane breathing mechanical defect mode is usually used to couple efficiently to a TE-like optical mode \cite{mayor2025high, Ren2020}. The defect unit cell at maximal displacement with corresponding mechanical band structure are shown in Fig. \ref{Fig1}d. To optimize the optomechanical coupling rate, the mechanical mode near the $\Gamma$ point is utilized, where the $k_\mathrm{m}=0$ mechanical wavevector satisfies the phase-matching condition with the standing optical cavity mode \cite{Amir2010}. The same choice that maximizes $g_{\mathrm{om}}$, however, places the mode at a band extremum where the mechanical group velocity vanishes, and in these 2D crystals the breathing band remains nearly flat across the entire Brillouin zone (Fig. \ref{Fig1}d). Slowly propagating phonons are easily localized: once fabrication disorder shifts the local band by more than its already-small dispersion, extended waveguide modes fragment into disconnected ones. Establishing a frequency-matched traveling-wave link to such a flat band is therefore impractical, and the component-matching strategy of Fig. \ref{Fig1}b offers no robust route to an electromechanical interface in two dimensions.

\begin{figure*}
\centering
\includegraphics[width=1\linewidth]{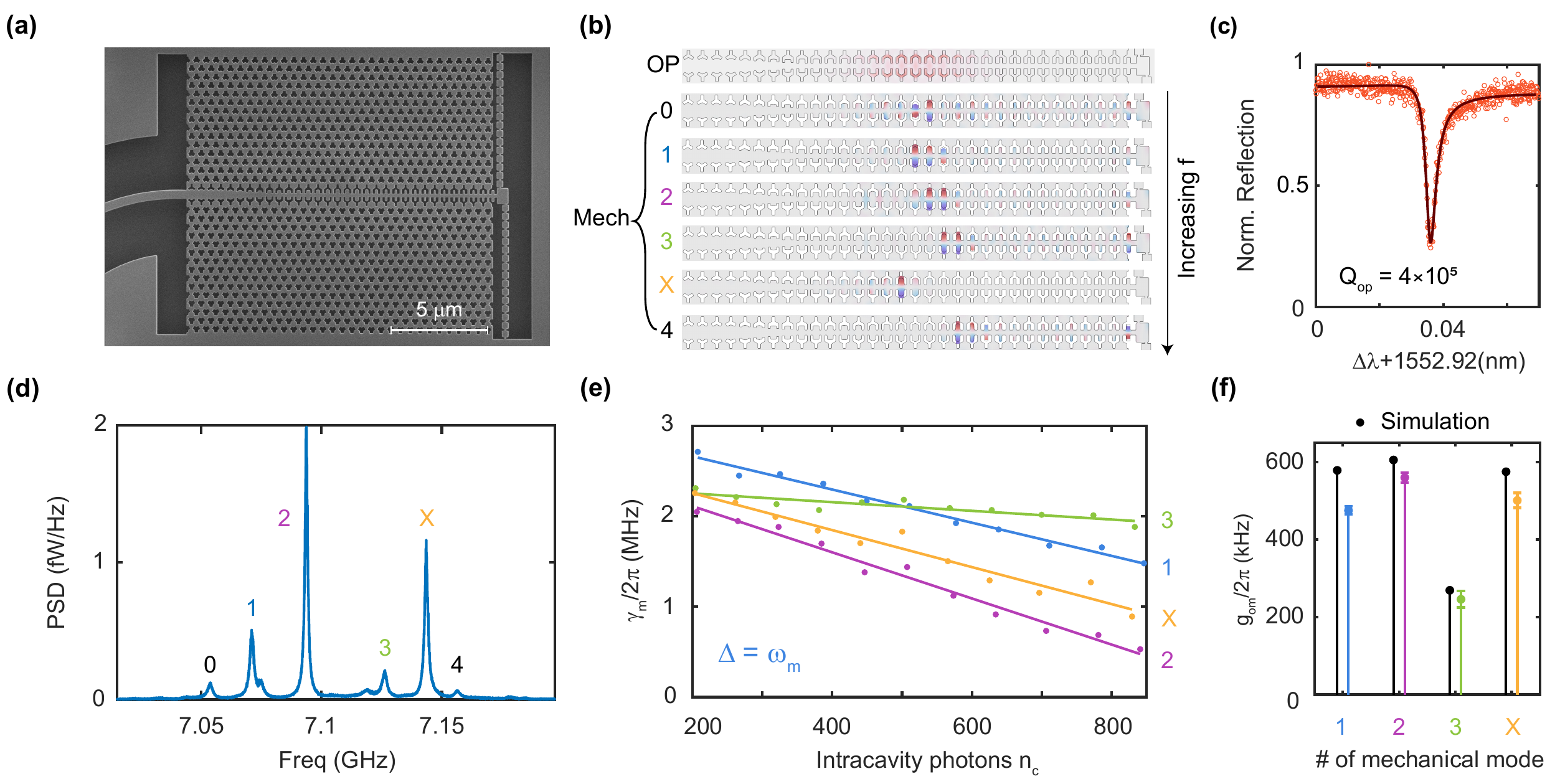}
\caption{\label{Fig3}
\textbf{Characterization of the collective mechanical modes from a 2D OMC-only device.}
\textbf{a}, Scanning electron microscope image of a 2D OMC-only device. The engineered $\Gamma$-potential is terminated by a silicon plate at the location where the LN block is designed to be transfer-printed in the full transducer architecture. 
\textbf{b}, Simulated mechanical modes confined by the engineered $\Gamma$-potential. Mode 4 arises from a local cavity formed between the input mirror and the central defect region. 
\textbf{c}, Measured optical reflection spectrum, showing an optical quality factor of $4\times10^5$.
\textbf{d}, Room-temperature thermomechanical spectrum of the device, showing a multi-mode mechanical response.
\textbf{e}, Optomechanical coupling rate measurement using the back-action method. $\gamma_{\mathrm{m}}$ is the mechanical linewidth. The back-action slope is proportional to $g_{\mathrm{om}}^2/\kappa_\mathrm{o}$ where $\kappa_{\mathrm{o}}$ is the optical cavity decay rate. Data are measured with a blue-detuned pump. 
\textbf{f}, Comparison between simulated and measured single-photon optomechanical coupling rates for the different mechanical modes. 
}
\end{figure*}

Here, we leverage the vanishing group velocity of the $\Gamma$-point breathing mode as a resource, as illustrated in Fig. \ref{Fig1}c. We structure the $\Gamma$-point frequency to follow a concave distribution across multiple unit cells, which we refer to as the $\Gamma$-potential. Frequency tuning is achieved by decreasing (increasing) the width of the central mode region in each unit cell, which shifts the entire band upward (downward). The mechanical bands in the $\Gamma$-potential region partially overlap in frequency between adjacent unit cells (Fig. \ref{Fig1}c). At a given mode frequency, real wavevectors exist across multiple unit cells, supporting spatially-extended mechanical motion. These collective mechanical modes are each bounded by a pair of unit cells excited at their stationary $\Gamma$-point. Due to the low group velocity at the $\Gamma$-point, the mode's mechanical energy concentrates on the boundaries and decays exponentially afterwards. As a result, small mode volumes are preserved, enabling large $g_\mathrm{om}$.

We verify this design strategy with finite-element simulations of a chain of optomechanical unit cells (Fig. \ref{Fig1}d) whose $\Gamma$-point frequencies follow a quadratic profile, obtaining the ladder of collective mechanical modes shown in Fig. \ref{Fig1}e. This construction is exactly analogous, within an envelope approximation, to the formation of bound states in a quantum harmonic oscillator \cite{scully1997quantum}. Near the $\Gamma$ point the breathing-mode band curves downward, $\omega(k_{\mathrm{m}},x) \approx \omega_{\Gamma}(x) - \alpha k_{\mathrm{m}}^{2}/2$, so the slowly varying envelope $\psi(x)$ of a collective mode at frequency $\omega$ satisfies a Schr\"odinger-like equation, $(\alpha/2)\,\partial_{x}^{2}\psi + \omega_{\Gamma}(x)\,\psi = \omega\,\psi$, describing a particle in a potential $\omega_{\Gamma}(x)$. A ladder of discrete eigenmodes descends from the center of the potential, with envelopes that, like excited oscillator states, concentrate near their classical turning points (Fig. \ref{Fig1}e and f). For a quadratic profile, the eigenfrequencies are equally spaced, with a splitting set by the geometric mean of the band curvature $\alpha$ and the curvature of $\omega_{\Gamma}(x)$; the simulated spectrum is also approximately linear (see Supplementary Materials).

\begin{figure*}
\centering
\includegraphics[width=1\linewidth]{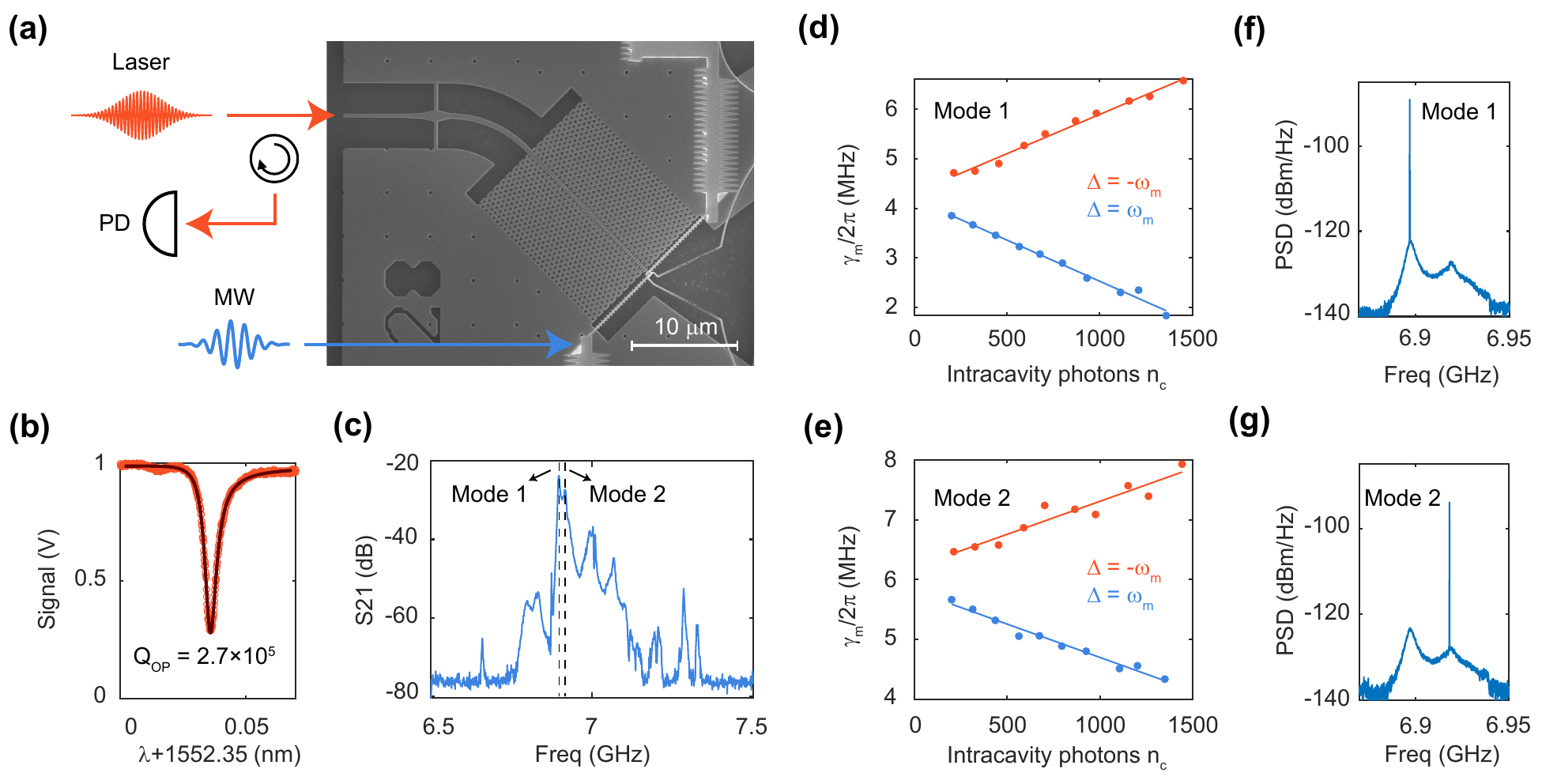}
\caption{\label{Fig4}
\textbf{Room-temperature characterization of a 2D piezo-optomechanical transducer.}
\textbf{a}, Scanning electron microscope image of a 2D piezo-optomechanical transducer, with schematics indicating the measurement setup. More details of the measurement are shown in Supplementary Materials. 
\textbf{b}, Measured optical reflection spectrum of the full transducer, showing an optical quality factor of $Q_{\mathrm{op}} =2.7\times10^5$.
\textbf{c}, Measured microwave-to-optical scattering response as a function of input microwave frequency. Two mechanical modes with relatively large scattering rates are indicated by dashed lines. 
\textbf{d,e}, Characterization of optomechanical coupling rates using both red- and blue-detuned optical pumps. We measure $g_\mathrm{om}/2\pi=$ 540 kHz for mode 1 and 446 kHz for mode 2. 
\textbf{f,g}, Characterization of electromechanical damping rates using blue-detuned optical pump. We extract $\gamma_\mathrm{em}/2\pi=$ 1.5 kHz for mode 1 and 4.6 kHz for mode 2. 
}
\end{figure*}

Importantly, these collective mechanical modes provide a natural electromechanical interface. By integrating piezoelectric material on one displacement antinode of the eigenmode, phonons can be electrically driven at the other antinode within the OMC defect region. Meanwhile, the design is also intrinsically less sensitive to fabrication disorder than a scheme based on matching several individual modes. Perturbations within the interior of the $\Gamma$-potential may slightly perturb the mode profile and eigenfrequency without destroying the collective mode itself. The architecture is most sensitive to disorder near the defect and piezoelectric region, where the local frequency sets the boundary of the $\Gamma$-potential. Such disorder can be tolerated as long as the induced frequency shifts remain smaller than the separation between adjacent $\Gamma$-point frequencies. This design therefore has the potential to provide a robust route to electromechanical coupling in a 2D piezo-optomechanical transducer.

\section*{Transducer modelling}
Motivated by the design principles introduced above, we implement a full piezo-optomechanical transducer on a heterogeneously integrated thin-film lithium niobate (LN) on released thin-film silicon (Si) platform. The device is formed by engineering a $\Gamma$-potential along the 2D OMC, and replacing the last unit cell with an LN block, as shown in Fig. \ref{Fig2}a. This geometry allows the collective mechanical mode supported by the $\Gamma$-potential to couple directly to the piezoelectric actuator while maintaining strong optical and mechanical confinement in the OMC cavity. To further improve the robustness of the electromechanical coupling, we utilize an asymmetric concave $\Gamma$-potential with larger frequency separation on the LN side to increase tolerance against fabrication disorder, as shown in Fig. \ref{Fig2}b. The asymmetry also enables a gentler optical taper for confining the optical mode. Additional unit cells between the OMC cavity and the LN suppress optical leakage.

Numerical simulations of the full transducer structure show a ladder of collective mechanical modes with an approximately linear eigenfrequency spectrum and mode spacings of $\sim 30~\mathrm{MHz}$, as shown in Fig. \ref{Fig2}c. The corresponding optical and mechanical mode profiles are displayed in Fig. \ref{Fig2}d with simulated radiation-limited optical quality factor of $7\times10^6$ and anchor-loss-limited mechanical quality factor of $1.4\times10^5$. We notice there is a localized non-collective mode labeled as \#X due to a surrounded mirror structure from the taper. Although this mode is with decent optomechanical coupling rate, the electromechanical coupling rate is vanishing due to the non-collecitve nature. From these simulated mode profiles, we extract a peak single-photon optomechanical coupling rate of $g_\mathrm{om}/2\pi = 595~\mathrm{kHz}$ (Fig. \ref{Fig2}e). The electromechanical damping rate for mode 1 is estimated to be $\gamma_\mathrm{em}/2\pi = 18~\mathrm{kHz}$ via a $50~\Omega$ probe with tens of nanometer tolerance on the LN block width, as shown in Fig. \ref{Fig2}f (see Supplementary Materials for details). Higher-order mechanical modes are more spatially compact, and therefore have reduced overlap with both the LN coupling region and the optical mode volume. As a result, both the optomechanical and electromechanical coupling rates decrease for higher-order modes. The engineered $\Gamma$-potential therefore increases the selectivity of collective mechanical modes by spectrally and spatially separating higher-order modes from the desired defect mode, minimizing their participation in the transduction process.

\section*{RT Measurement - OMC only}
To experimentally validate the design before integrating the piezoelectric interface, we fabricated a 2D OMC-only device, in which the engineered $\Gamma$-potential is terminated by a silicon plate at the desired LN position. Phononic shields connect the plate to the substrate, suppressing mechanical radiation into the environment, and also providing structural support for electrodes to interface with the LN in subsequent designs. A scanning electron microscope image of a nominal device is shown in Fig. \ref{Fig3}a with the simulated optical and mechanical eigenmode profiles shown in Fig. \ref{Fig3}b. The measured optical spectrum, shown in Fig. \ref{Fig3}c, exhibits an optical cavity resonance with a loaded quality factor of $Q_{\mathrm{op}} = 4\times10^5$, confirming that the engineered 2D OMC maintains strong optical confinement. 

We then characterize the mechanical modes of the 2D OMC cavity using thermal spectroscopy. The power spectral density (PSD) of the device at room temperature is shown in Fig. \ref{Fig3}d. Different modes exhibit different thermal power spectral densities, with peak height scaling in proportion to $g_{\mathrm{om}}^2/\gamma_\mathrm{m}$. To obtain the optomechanical coupling rate of each mechanical mode, we perform optomechanical backaction measurements \cite{Kippenberg2008} by sweeping the power of a blue-detuned pump, and measuring the change in the linewidth of the mechanical mode. Fig. \ref{Fig3}e shows the measurement from mode 1 to mode 4 (color-coded for reference); mode 0 and mode 5 are omitted due to their low optomechanical coupling rates. From these measurements, we extract the single-photon optomechanical coupling rates and summarize the results in Fig. \ref{Fig3}f. The measured coupling rates agree well with finite-element simulations, validating the understanding of mechanical modes arising from the engineered $\Gamma$-potential prior to LN integration.

\section*{RT Measurement - full transducer}
Having validated the key features of the engineered collective modes in the pure 2D OMC devices, we next fabricate full piezo-optomechanical transducers with a transfer-printed LN block and patterned electrodes for the microwave interface \cite{Jiang2023}. A simplified measurement schematic is shown in Fig. \ref{Fig4}a. Details of the fabrication process and experimental setup are provided in the Supplementary Materials.

We first characterize the optical response of the 2D transducer. The measured optical quality factor is $Q_{\mathrm{op}} = 2.7\times10^5$ (Fig. \ref{Fig4}b), comparable to the value obtained from the OMC-only device. This indicates that the LN transfer-print process and subsequent electrode fabrication do not substantially degrade the optical quality factor of the silicon OMC cavity. We then measure the microwave-to-optical scattering response using a blue-detuned optical pump with a $\sim500~\mu\mathrm{W}$ pump power (Fig. \ref{Fig4}c). Two mechanical modes with relatively large scattering rates are observed and labeled as mode 1 and 2.

We next characterize the optomechanical coupling rates of these modes using both red- and blue-detuned optical pumps, as shown in Fig. \ref{Fig4}d and \ref{Fig4}e. For modes 1 and 2, we extract single-photon optomechanical coupling rates of $g_{\mathrm{om}}/2\pi = 540~\mathrm{kHz}$ and $446~\mathrm{kHz}$ respectively, consistent with the expected large optical-mechanical overlap of the collective modes. The measured backaction-free room-temperature mechanical linewidths are $4.2~\mathrm{MHz}$ and $6.0~\mathrm{MHz}$. We attribute these linewidths primarily to room-temperature mechanical damping and expect them to become substantially narrower at cryogenic temperatures\cite{Chant2012}.

We further characterize the electromechanical coupling by applying a single coherent microwave tone at the mode frequency, as shown in Fig. \ref{Fig4}f and \ref{Fig4}g. The microwave drive produces a narrow converted mechanical response on top of the thermally populated mechanical spectrum. By normalizing to the room-temperature thermomechanical emission, $k_{\mathrm{B}}T/\hbar\omega_\mathrm{m}$, we infer the converted mechanical occupation from the applied microwave drive $P_{\mathrm{e,in}}=-61~\mathrm{dBm}$. From this calibration, we extract electromechanical damping rates at $1.5~\mathrm{kHz}$ and $4.6~\mathrm{kHz}$ for modes 1 and 2, respectively. The key parameters of these two modes are summarized in Supplementary Materials. We note that the calibrated direct electromechanical conversion efficiency of $6.2\times10^{-4}$ are obtained in the absence of a high-impedance microwave resonator. This efficiency will be enhanced by a frequency-matched microwave resonator at cryogenic temperatures.

\begin{figure}
\centering
\includegraphics[width=1\linewidth]{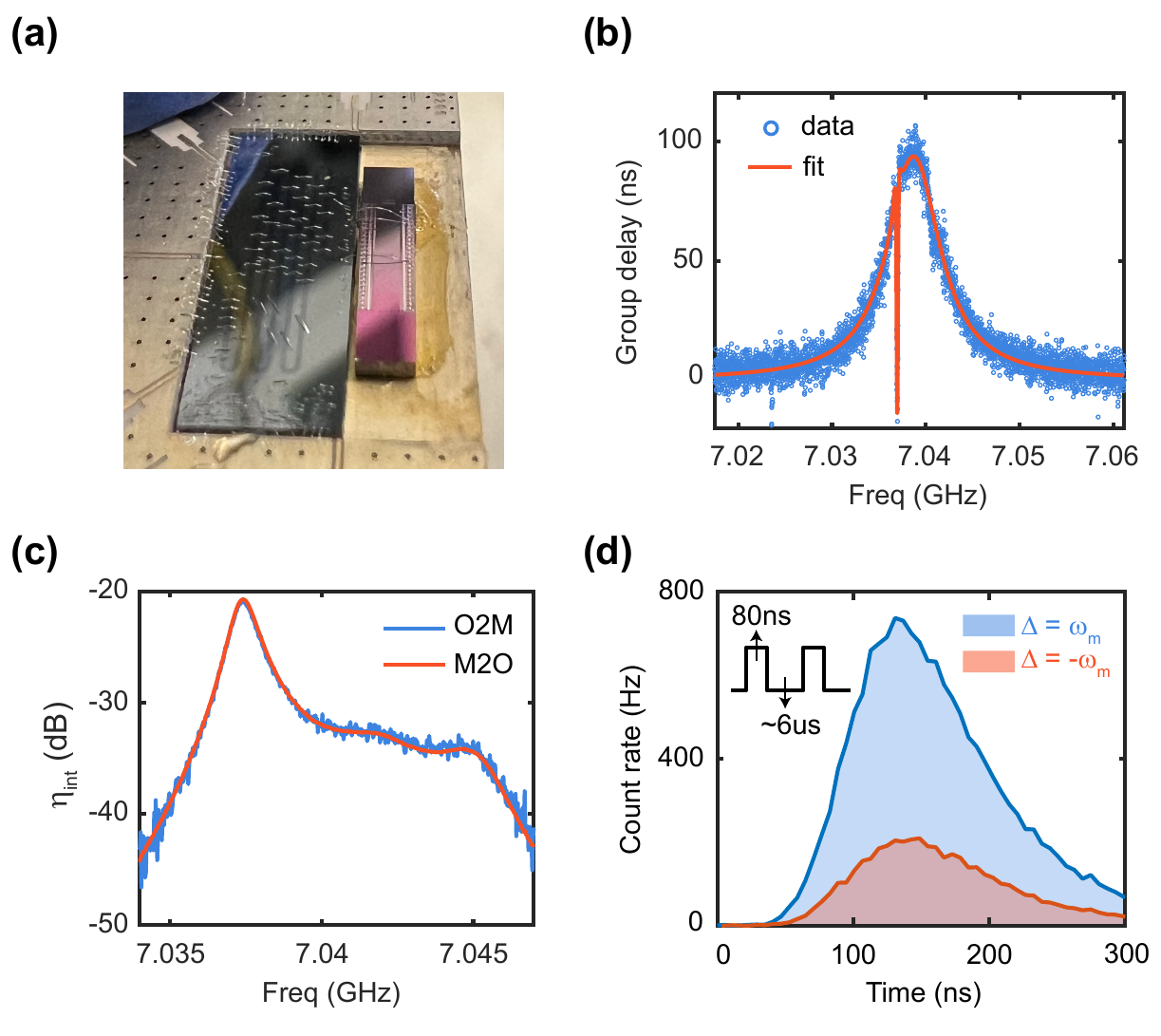}
\caption{\label{Fig5}
\textbf{Cryogenic characterization of a 2D piezo-optomechanical transducer.}
\textbf{a}, Image of the packaged device. The transducer chip is wire-bonded to a tunable multi-mode microwave resonator.
\textbf{b}, Measured microwave group delay, showing an over-coupled microwave resonator coupled to a mechanical mode. A fit based on coupled-mode theory yields an electromechanical coupling rate of $g_{\mathrm{em}}/2\pi = 0.17~\mathrm{MHz}$ and $\gamma_{\mathrm{m}}/2\pi = 310~\mathrm{kHz}$.
\textbf{c}, Calibrated internal transduction efficiency measured at a continuous-wave optical pump power of $500~\mu\mathrm{W}$ for both optical-to-microwave (O2M) and microwave-to-optical (M2O) conversion.
\textbf{d}, Time traces of sideband photons generated by blue- and red-detuned pump pulses. The blue-detuned pulse produces photon-phonon pairs with a scattering probability of $5\%$ using 80 ns pulse length. At a pulse repetition rate of $168~\mathrm{kHz}$, this corresponds to a photon-phonon pair generation rate of $8.4~\mathrm{kHz}$. From the asymmetry between the blue- and red-detuned sideband count rates, we extract a mechanical phonon occupation of $n_{\mathrm{m}}=0.41$.
}
\end{figure}

\section*{Cryogenic Measurement - Full Transducer}
To further demonstrate the capacity under conditions relevant for quantum transduction, we wire-bond the same transducer chip to a tunable multi-mode microwave resonator \cite{Jiang2023} and characterize the system at 10 mK, as shown in Fig. \ref{Fig5}a. By tuning the microwave resonator with an external magnetic field, we measure the microwave group delay and observe a coupled electromechanical response consisting of an over-coupled microwave resonator and a mechanical mode (Fig. \ref{Fig5}b). The corresponding amplitude and phase responses are provided in the Supplementary Materials. Fitting the response with a coupled-mode model yields an electromechanical coupling rate of $g_{\mathrm{em}}/2\pi = 0.17~\mathrm{MHz}$ and $\gamma_{\mathrm{m}}/2\pi = 310~\mathrm{kHz}$, which is expected to be lower than that of prior transducers, primarily due to the parasitic capacitance from the wire-bonds and the large mode volume of the multi-mode microwave resonator. A more optimized structure with a smaller co-integrated or bump bonded microwave resonator would lead to larger electromechanical coupling \cite{Malik2023}.

We next demonstrate bidirectional microwave-optical transduction by measuring both microwave-to-optical (M2O) and optical-to-microwave (O2M) conversion. At a continuous-wave optical pump power of $500~\mu\mathrm{W}$, we calibrate an internal conversion efficiency of up to $0.85\%$, as shown in Fig. \ref{Fig5}c. 
In addition to coherent conversion, we further characterize the thermal response of the transducer with a pulsed optical pump. With a scattering probability of $5\%$ at 80 ns pulse length and a pulse repetition rate of $168~\mathrm{kHz}$, the device generates photon-phonon pairs at a rate of $8.4~\mathrm{kHz}$, as shown in Fig. \ref{Fig5}d. By comparing the sideband photon counts generated by blue- and red-detuned pump pulses, we extract a transient mechanical phonon occupation of $n_{\mathrm{m}}=0.41$. This sub-unity phonon occupation indicates that the mechanical mode is near its quantum ground state during the pulsed measurement. Detailed thermal behavior will be explored in future studies. Together, these cryogenic measurements demonstrate that the engineered collective mechanical modes enable robust electromechanical integration while preserving the advantages of 2D OMC transducers. The observed bidirectional transduction, photon-phonon pair generation at sub-unity mechanical occupation establish a clear path toward high-efficiency and low-noise 2D piezo-optomechanical transducers.

\section*{Conclusion}
In this work, we introduce a new design strategy for realizing robust electromechanical coupling in 2D OMC-based piezo-optomechanical transducers. By engineering the band structure of unit cells, we directly design a collective mechanical mode through a spatially varying $\Gamma$-potential. This effective potential supports extended mechanical modes that provide direct access to a piezoelectric interface while remaining robust against moderate fabrication disorder. We verify the design strategy experimentally, by fabricated OMC-only devices to confirm the multi-mode structures and LN-integrated 2D transducers to confirm the electromechanical interface. From these measurements, we extract an electromechanical damping rate of $\gamma_{\mathrm{em}}/2\pi=4.6~\mathrm{kHz}$ and a direct electromechanical conversion efficiency of $\eta_{\mathrm{em}} = 6.2\times10^{-4}$ without microwave resonator enhancement. Cryogenic measurements are performed with a wire-bonded multi-mode microwave resonator. A sub-MHz electromechanical coupling rate is observed, similar to the coupling strength from 1D transducer measurements \cite{Jiang2023}. These results place the 2D transducer on a comparable footing with previously demonstrated 1D piezo-optomechanical transducers, while retaining the thermal advantages expected from the 2D OMC platform. 

Future work will focus on replacing the electrode material of the transducer from aluminum electrodes to niobium, which would be compatible with an acid clean prior to cryogenic measurements that could help cleaning the surface oxide and reduce surface two-level-system (TLS) density. Concurrently, integrating the 2D OMC transducer with a high-impedance superconducting microwave resonator \cite{Malik2023, Koolstra2025} would be valuable for attaining high efficiency operation as well as microwave cooling. The demonstrated wirebond-based microwave interfaces introduced significant parasitic capacitances, which limited the achievable electromechanical coupling rate. Reducing the separation distance between the transducer and the microwave resonator via flip-chip direct galvanic integration \cite{Malik2023} should alleviate this issue. With these future integrations in place, the full transducer would be characterized at cryogenic temperatures, where we expect improved mechanical quality factor, improved thermal response both from a clean surface and radiative cooling from the microwave channel \cite{Zhao2025,Xu2020}. We anticipate that this architecture can reach electromechanical coupling rates in the $10~\mathrm{MHz}$ range and substantially improve the added-noise performance compared with current 1D OMC transducers. These results pave the way toward a high-efficiency, low-noise and high-repetition rate microwave-to-optical transducer for entangling remote superconducting quantum processors.

\begin{acknowledgments}
We thank Gitanjali Multani, Yizhi Luo and Yun Zhao for helpful discussion and technical assistance. This work was primarily supported by the US Army Research Office (ARO)/ Laboratory for Physical Sciences (LPS) Modular Quantum Gates (ModQ) program (Grant No. W911NF-23-1-0254). Some of this work was funded by the US Department of Energy through grant no. DE-AC02-76SF00515 and via the Q-NEXT Center, and the Gordon and Betty Moore Foundation, Grant 12214. We also thank Amazon Web Services Inc. for their financial support. S.G. acknowledges support by the Swiss National Science Foundation (SNSF) [225443] and the Knut and Alice Wallenberg foundation [KAW 2021-
0341]. Device fabrication was performed at the Stanford Nano Shared Facilities (SNSF) and the Stanford Nanofabrication Facility (SNF), supported by NSF award ECCS-2026822. 

\textbf{Author contributions:} T.X. and N.O. contributed equally to this work. T.X. and A.H.S.-N. conceived the idea. 
T.X. and N.O. designed the device with assistance from S.M., F.M.M., S.G. and A.G.P.. 
T.X. and N.O. fabricated the device with assistance by L.W., S.M., F.M.M. and S.G..
F.M.M, S.G, O.A.H, W.J. and S.M. developed the fabrication process.
S.M. designed and fabricated the multi-mode microwave resonator. 
T.X., N.O. and L.W. measured the device with assistance from S.M., F.M.M. and S.G..
T.X. and N.O. performed the data analysis. 
T.X., N.O and A.H.S.-N. wrote the manuscript with input from all authors. A.H.S.-N. supervised the project.

\textbf{Data and materials availability:} The data that support the findings of this study are available from the corresponding author upon reasonable request.

\textbf{Competing interests:} A.H.S.-N. has been an Amazon Scholar and consultant for Ely Sensor Technologies during this work. The other authors declare no competing interests.

\end{acknowledgments}

\bibliographystyle{naturemag}
\bibliography{ref}

\clearpage

\renewcommand{\thefigure}{S\arabic{figure}}
\renewcommand{\thetable}{S\arabic{table}}
\renewcommand{\theequation}{S\arabic{equation}}

\onecolumngrid
\begin{center}
    {\large Supplementary Materials\par}
\end{center}
\setcounter{figure}{0}

\section{Theoretical model of piezo-optomechanical transducers.}

\subsection{Definitions.}

\begin{center}
\begin{tabular}{|c|l|}
\hline
$\hat{a}$ & Annihilation operator for optical cavity mode, with angular frequency $\omega_\mathrm{o}$.\\
$\hat{b}$ & Annihilation operator for mechanical cavity mode, with angular frequency $\omega_\mathrm{m}$.\\
$\hat{c}$ & Annihilation operator for microwave cavity mode, with angular frequency $\omega_\mathrm{e}$.\\
$\kappa_{\mathrm{ex}}/ 2\pi$ & External damping rate of optical cavity to input waveguide. \\
$\kappa/ 2\pi$ & Total damping rate of the optical cavity. \\

$\gamma_{\mathrm{ex}}/ 2\pi$ & External damping rate of mechanical cavity to the phonon bath. \\

$\gamma_\mathrm{m}/ 2\pi$ & Total damping rate of the mechanical cavity. \\

$\kappa_{\mathrm{e, ex}}/ 2\pi$ & External damping rate of microwave cavity to input waveguide. \\

$\kappa_\mathrm{e} / 2\pi$ & Total damping rate of the microwave cavity. \\
$\gamma_{\mathrm{om}} / 2\pi$ & Optomechanical damping rate of mechanical mode. \\
$\gamma_{\mathrm{em}} / 2\pi$ & Electromechanical damping rate of mechanical mode. \\
$g_{\mathrm{om}} / 2\pi$ & Single-photon optomechanical coupling rate between optical and mechanical cavity modes. \\
$g_{\mathrm{em}} / 2\pi$ & Electromechanical coupling rate between microwave and mechanical cavity modes. \\
$n_{\mathrm{cav}}$ & Circulating photon number in the optical cavity. \\

\hline
\end{tabular}
\end{center}

In this section we derive the primary metrics that characterize the piezo-optomechanical transducer. For a background on the theory of piezo-optomechanical transduction, we refer the reader to Refs. \citeSI{Jiangt2022, han2021}
.

\subsection{Optomechanical and electromechanical damping rate.}

The piezo-optomechanical transducer can be modeled as a three-port device with one microwave, one mechanical, and one optical port. Only the microwave or optical ports may be directly probed, while the mechanical port is coupled to a mechanical bath. Under red-sideband optical pumping in the resolved-sideband regime, the linearized optomechanical interaction Hamiltonian follows a beamsplitter form ($\hat{a}^\dagger\hat{b} + \hat{a}\hat{b}^\dagger$). The Heisenberg-Langevin equations in the frequency domain are given by

\begin{equation} \label{opt_lang}
-i\omega\hat{a}(\omega) = \left(i\Delta - \frac{\kappa}{2}\right)\hat{a}(\omega)- i g_{\mathrm{om}}\sqrt{n_\mathrm{cav}}{\hat{b}(\omega)} - \sqrt{\kappa_{\mathrm{ex}}} \hat{a}_{\mathrm{in}}(\omega)
\end{equation}
\begin{equation}
-i\omega{\hat{b} (\omega)} = - \left(i\omega_\mathrm{m} + \frac{\gamma_\mathrm{m}}{2}\right){\hat{b}(\omega)} - ig_{\mathrm{om}}\sqrt{n_\mathrm{cav}} \hat{a}(\omega) - i g_\mathrm{em} \hat{c}(\omega) - \sqrt{\gamma_\mathrm{m}} \hat{b}_\mathrm{in}(\omega)
\end{equation}

\begin{equation}
-i\omega\hat{c}(\omega) = - \left(i\omega_\mathrm{e} + \frac{\kappa_\mathrm{e}}{2}\right){\hat{c}(\omega)} - i g_\mathrm{em} \hat{b}(\omega) - \sqrt{\kappa_{\mathrm{e}, \mathrm{ex}}} \hat{c}_{\mathrm{in}}(\omega)
\end{equation}

where $\Delta = \omega_\mathrm{L} - \omega_\mathrm{o} = -\omega_\mathrm{m}$ corresponds to the red sideband. The associated output boundary conditions for optical and microwave modes are given by $\hat{a}_\mathrm{out} = \hat{a}_\mathrm{in} + \sqrt{\kappa_{\mathrm{ex}}} \hat{a}$ and 
$\hat{c}_\mathrm{out} = \hat{c}_\mathrm{in} + \sqrt{\kappa_{\mathrm{e,ex}}} \hat{c}$ respectively. For optomechanical crystal-based transducers, it is often valid to adiabatically eliminate the optical and mechanical modes since they generally operate in the regime where $\kappa_\mathrm{ex} \gg g_\mathrm{om}\sqrt{n_\mathrm{cav}}$, $\kappa_\mathrm{e, ex} \gg g_\mathrm{em}$.

Following from eq. \ref{opt_lang} and neglecting the optical noise inputs, the effective coupling between the mechanical and optical fields is given by

\begin{equation}
    \hat{a} = -\frac{2ig_\mathrm{om}\sqrt{n_\mathrm{cav}}}{\kappa}\hat{b} \implies \hat{a}_\mathrm{out} = -\sqrt{\frac{\kappa_\mathrm{ex}}{\kappa}}\frac{2ig_\mathrm{om}\sqrt{n_\mathrm{cav}}}{\sqrt{\kappa}} \hat{b}
\end{equation}

And using the adiabatic elimination assumption, the phonon population thus decays into the optical port at an effective rate given by 

\begin{equation}
\gamma_\mathrm{om} = \left|\frac{2ig_\mathrm{om}\sqrt{n_\mathrm{cav}}}{\sqrt{\kappa}}\right|^2 = \frac{4g_\mathrm{om}^2n_\mathrm{cav}}{\kappa}    
\end{equation}

with collection efficiency $\eta_\mathrm{o} = \kappa_\mathrm{ex}/\kappa$. A similar treatment for the effective mechanical-microwave coupling gives $\gamma_\mathrm{em} = {4g_\mathrm{em}^2}/{\kappa_\mathrm{e}}$ with collection efficiency $\eta_\mathrm{e} = {\kappa_{\mathrm{e, ex}}}/{\kappa_\mathrm{e}}$ .

The optomechanical and electromechanical cooperativities in terms of their respective damping rates are then given by 

\begin{equation}
C_\mathrm{om} = \frac{4 g_\mathrm{om}^2n_\mathrm{cav}}{{\kappa \gamma_\mathrm{m}}} = \frac{\gamma_\mathrm{om}}{\gamma_\mathrm{m}}
\end{equation}
\begin{equation}
C_\mathrm{em} = \frac{4 g_\mathrm{em}^2}{\kappa_\mathrm{e} \gamma_\mathrm{m}} = \frac{\gamma_\mathrm{em}}{\gamma_\mathrm{m}}.
\end{equation}

\subsection{Calibration of room-temperature electromechanical conversion efficiency.}

We consider the case where the optomechanical transducer's microwave channel is directly coupled to the mechanical mode at the electromechanical damping rate $\gamma_\mathrm{em}$, without resonant enhancement. Neglecting the noise input to the mechanical mode, the equations of motion are then given by

\begin{equation} 
-i\omega\hat{a}(\omega) = \left(i\Delta - \frac{\kappa}{2}\right)\hat{a}(\omega)- i g_\mathrm{om}\sqrt{n_{\mathrm{cav}}}{\hat{b}(\omega)} - \sqrt{\kappa_{\mathrm{ex}}} \hat{a}_{\mathrm{in}}(\omega)
\end{equation}
\begin{equation}\label{direct_mech_couple}
-i\omega{\hat{b} (\omega)} = - \left(i\omega_\mathrm{m} + \frac{\gamma_\mathrm{m}}{2}\right){\hat{b}(\omega)} - ig_\mathrm{om}\sqrt{n_{\mathrm{cav}}}\hat{a}(\omega) - \sqrt{\gamma_\mathrm{em}} \hat{c}_\mathrm{in}(\omega)
\end{equation}

Again, we let $A = -i(\Delta + \omega)+\kappa/2, B = i(\omega_\mathrm{m} - \omega)+\gamma_\mathrm{m}/2$. Using the matrix inversion solution to this coupled set of equations, we have

\begin{equation}\label{no_microwave_cavity_transduction}
\frac{1}{1 + g_\mathrm{om}^2n_\mathrm{cav}/AB}
\begin{pmatrix}
1 & -ig_\mathrm{om}\sqrt{n_{\mathrm{cav}}}/A \\
-ig_\mathrm{om}\sqrt{n_{\mathrm{cav}}}/B & 1\\
\end{pmatrix}
\begin{pmatrix}
-\sqrt{\kappa_\mathrm{ex}}a_\mathrm{in}/A \\
-\sqrt{\gamma_\mathrm{em}}c_\mathrm{in}/B \\
\end{pmatrix} =
\begin{pmatrix}
a \\
b
\end{pmatrix}
\end{equation}

Reading off from this, the mechanical field amplitude is given by

\begin{equation}
    S_\mathrm{bc} = \frac{-\sqrt{\gamma_\mathrm{em}}}{i(\omega_\mathrm{m} - \omega) + \gamma_\mathrm{m}/2 + \frac{g_\mathrm{om}^2n_\mathrm{cav}}{-i(\Delta + \omega) + \kappa/2}}
\end{equation}

At the red sideband frequency $\omega = -\Delta = \omega_\mathrm{e}$ we can relate the number of coherent phonons to the microwave flux by the following

\begin{equation}
    |S_\mathrm{bc}|^2 |\hat{c}_\mathrm{in}|^2 = n_\mathrm{m,coh} = \frac{\gamma_\mathrm{em}\dot{N}_\mathrm{e}}{(\omega_\mathrm{e} - \omega_\mathrm{m})^2 + (\gamma_\mathrm{tot}/2)^2}
\end{equation}

Where $\gamma_\mathrm{tot} = \gamma_\mathrm{m} + \gamma_\mathrm{om}$, and the input microwave photon flux $\dot{N}_\mathrm{e} = |\hat{c}_\mathrm{in}|^2 = P_\mathrm{e, in} / \hbar \omega_\mathrm{e}$. On resonance at $\omega_\mathrm{e} = \omega_\mathrm{m}$ the effective microwave to mechanical coupling rate is given by

\begin{equation}
    \gamma_{\mathrm{em}} = \frac{n_\mathrm{m,coh}\gamma_\mathrm{tot}^2}{4 \dot N_\mathrm{e}}
\end{equation}

We apply a coherent single microwave tone to the microwave port of the transducer. Using an optical probe at the appropriate detuning, we read out the power scattered into the optical sideband generated by the thermal phonon population, by collecting photons on a high speed photodiode. Integrating across the broad thermal Lorentzian corresponding to a single mechanical mode gives us its $P_\mathrm{therm}$, as shown in Fig \ref{Fig4}f and \ref{Fig4}g of the main text. Assuming the same optomechanical transduction gain for both incoherent and coherent phonon populations, we deduce the coherent phonon population given by 

\begin{equation}
    n_\mathrm{m,coh} = P_\mathrm{coh} \cdot \frac{n_\mathrm{therm}}{P_{\mathrm{therm}}}
\end{equation}

where we estimate the thermal phonon occupation at room temperature $T = 295 \mathrm{K}$ to be $n_\mathrm{therm} = k_\mathrm{B}T/\hbar \omega_\mathrm{m} \approx 900$.

The microwave-to-mechanics conversion efficiency $\eta_\mathrm{em}$ is then given by the electromechanical cooperativity

\begin{equation}
    \eta_\mathrm{em} = C_\mathrm{em} =\frac{\gamma_\mathrm{em}}{\gamma_\mathrm{m}}.
\end{equation}

The value obtained for $\gamma_\mathrm{em}$ represents the coupling from the microwave \textit{channel} to the mechanics. It is possible to transform this value to estimate a coupling rate between a microwave \textit{resonator} to the mechanical mode. This is given by

\begin{equation}
    g_\mathrm{em} = \frac{1}{2}\sqrt{\gamma_\mathrm{em}\omega_\mathrm{m}}\sqrt{\frac{Z_\mathrm{c}}{Z_0}}
    \label{SI_gem}
\end{equation}

where $Z_\mathrm{c}$ is the characteristic impedance of the microwave resonator (chosen to be 700 $\Omega$ in our experiment), and $Z_0$ is the characteristic impedance of the 50 $\Omega$ probe.

\subsection{Characterization of $g_\mathrm{om}$ at room temperature.}

In this analysis, we neglect the microwave port, and use a blue-sideband optical pump ($\Delta = +\omega_\mathrm{m}$), which produces a two-mode squeezing optomechanical interaction Hamiltonian ($\hat{a}^\dagger\hat{b}^\dagger + \hat{a}\hat{b}$). The Heisenberg-Langevin equations of motion are then given by

\begin{equation}
    -i\omega\hat{a}(\omega) = \left(i\Delta - \frac{\kappa}{2}\right)\hat{a}(\omega) -ig_\mathrm{om}\sqrt{n_\mathrm{cav}}\hat{b}^\dagger(\omega) - \sqrt{\kappa_\mathrm{ex}} \hat{a}_\mathrm{in}(\omega)
\end{equation}

\begin{equation}
    -i\omega\hat{b}(\omega) = -\left(i\omega_\mathrm{m} + \frac{\gamma_\mathrm{m}}{2}\right)\hat{b}(\omega) -ig_\mathrm{om}\sqrt{n_\mathrm{cav}}\hat{a}^\dagger(\omega) - \sqrt{\gamma_\mathrm{m}} \hat{b}_\mathrm{in}(\omega)
\end{equation}

which, after rearranging, yields

\begin{equation}
    -i\omega\hat{b}(\omega) = -\left(i\omega_\mathrm{m} + \frac{\gamma_\mathrm{m}}{2}\right)\hat{b}(\omega) +\frac{g_\mathrm{om}^2{n_\mathrm{cav}}\hat{b}(\omega)}{-i(\Delta + \omega) + \kappa/2} + \frac{ig_\mathrm{om}\sqrt{n_\mathrm{cav}}\sqrt{\kappa_\mathrm{ex}}\hat{a}_\mathrm{in}^\dagger}{{-i(\Delta + \omega) + \kappa/2}} - \sqrt{\gamma_\mathrm{m}} \hat{b}_\mathrm{in}(\omega)
\end{equation}

When near resonance, $\omega \approx -\Delta \approx -\omega_\mathrm{m}$ and the above simplifies to

\begin{equation}
    -i\omega\hat{b}(\omega) = -\left(i\omega_\mathrm{m} + \frac{\gamma_\mathrm{m} - \gamma_\mathrm{om}}{2}\right)\hat{b}(\omega) + i\sqrt{\gamma_\mathrm{om}}\sqrt{\frac{\kappa_\mathrm{ex}}{\kappa}}\hat{a}_\mathrm{in}^\dagger - \sqrt{\gamma_\mathrm{m}} \hat{b}_\mathrm{in}(\omega).
\end{equation}

Hence, the effective linewidth of the mechanical mode ($\gamma_\mathrm{tot} = \gamma_\mathrm{m}-\gamma_\mathrm{om}$) decreases linearly in $\gamma_\mathrm{om} \propto n_\mathrm{cav}$ when driving on the blue sideband. On the red sideband, the converse happens, with the effective mechanical linewidth ($\gamma_\mathrm{tot} = \gamma_\mathrm{m}+\gamma_\mathrm{om}$) broadening in $\gamma_\mathrm{om}$. Changing the optical input power alters $n_\mathrm{cav}$; this results in a corresponding change in $\gamma_\mathrm{om}$ which we can obtain via successive Lorentzian fits. By performing a linear fit on the plot of linewidth versus intracavity photon number, we find the gradient and calculate the single-photon $g_\mathrm{om}$. The back-action free mechanical linewidth is obtained from the intercept of the best-fit line at the $n_{\mathrm{cav}} = 0$ axis.

\section{Device design}

\begin{figure*}
\centering
\includegraphics[width=1\linewidth]{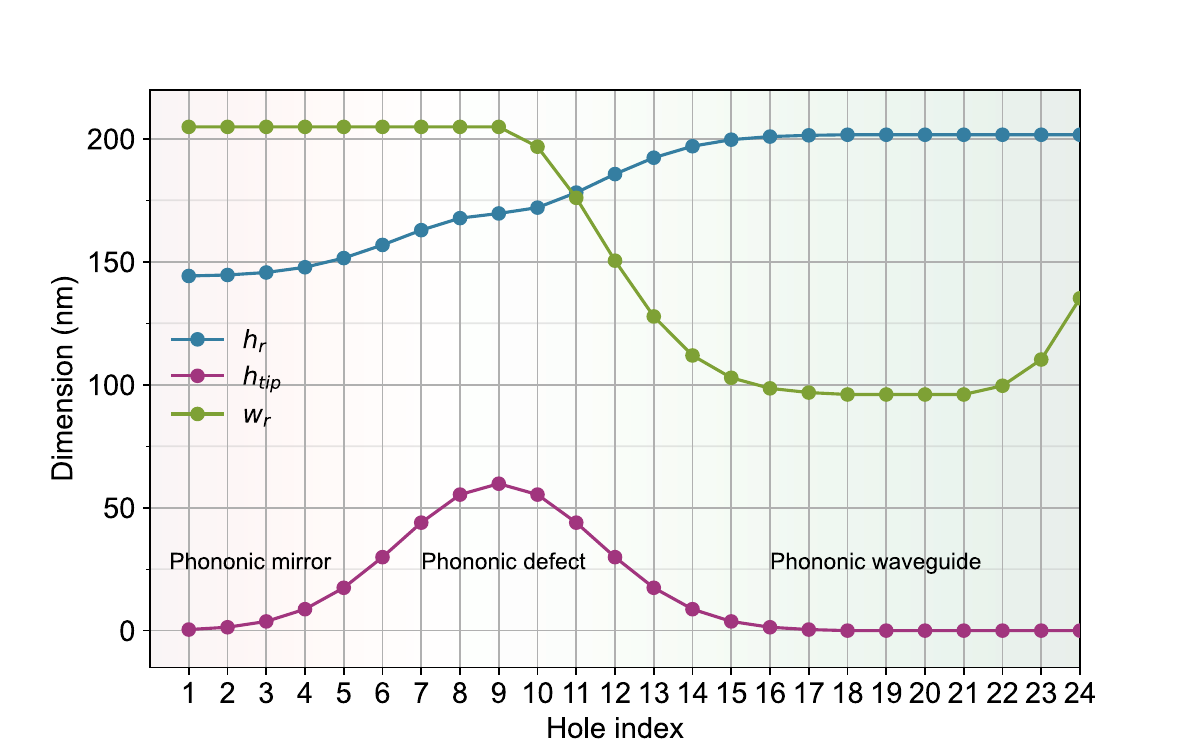}
\caption{\label{FigTaper}
{\textbf{Device geometry.} Dimensions of the unit cell are varied across a 2D OMC. Index 24 corresponds to the unit cell closest to the LN. Index 1 is adjacent to the optical input waveguide.}
}
\end{figure*}

The $\Gamma$-point mechanical frequency is governed by the width of the central silicon 'blade' $\mathrm{w}_{\mathrm{r}}$, the height of the straightedge on the silicon blade $\mathrm{h}_\mathrm{r}$, and the height of the blade's triangular region $\mathrm{h}_{\mathrm{tip}}$. These dimensions are shown in the Fig. \ref{Fig2}a inset of the main text; the taper shown in Fig. \ref{FigTaper} produces the $\Gamma$-point potential shown in Fig. \ref{Fig2}b.

\begin{figure*}
\centering
\includegraphics[width=1\linewidth]{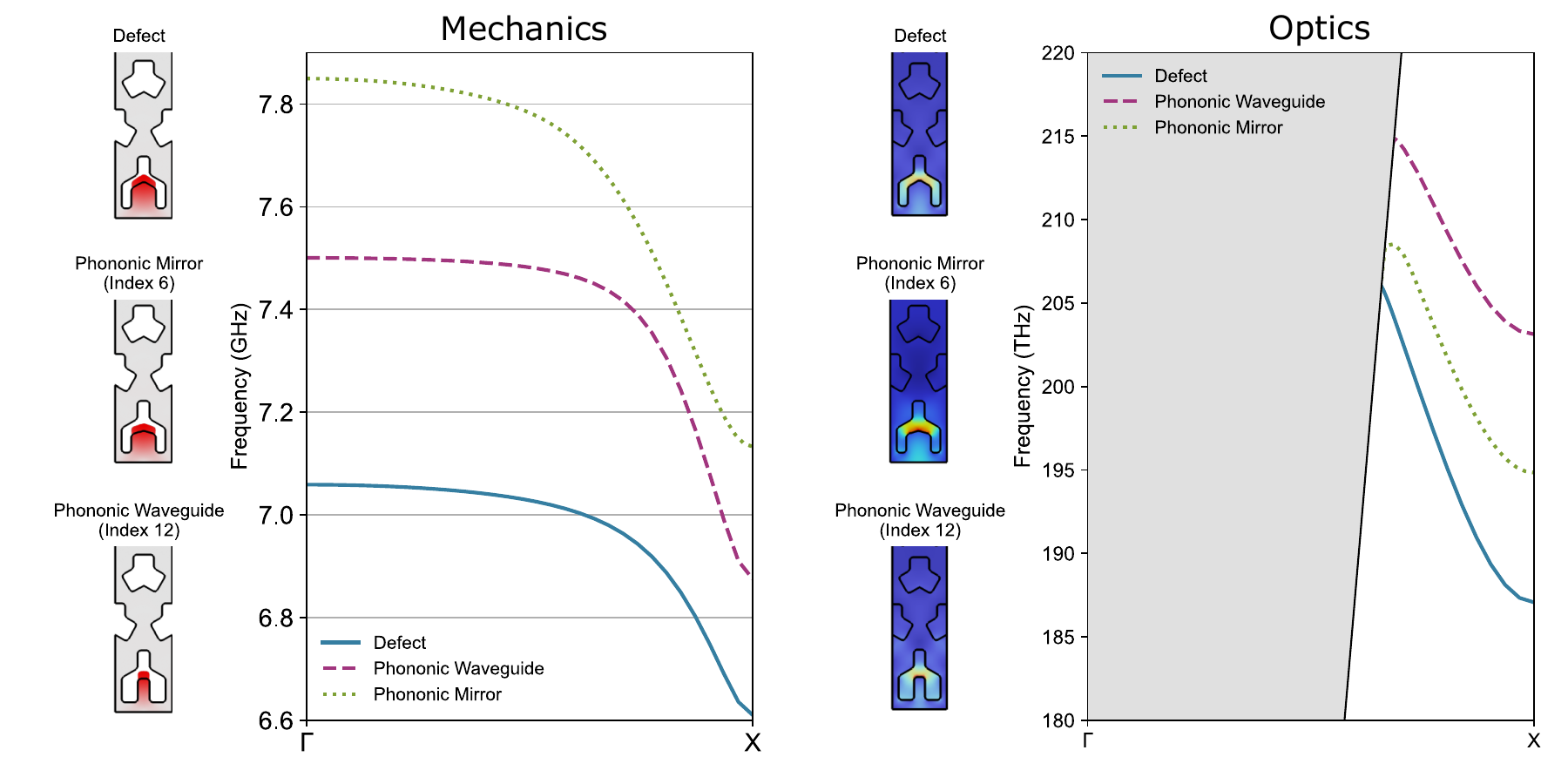}
\caption{\label{FigBand}
\textbf{Band structures of unit cells.} The mechanical and optical bands of interest are shown for representative unit cells in the defect, phononic waveguide, and phononic mirror regions of the OMC. These demonstrate the effect of varying dimensions in the unit cell structure. The shaded gray are in the optical plot corresponds to the continuum of unguided modes; its black border corresponds to the light line.
}
\end{figure*}

In an OMC cavity, the geometry of the thin-film silicon membrane governs both the optical and mechanical properties. As such, we also engineer an input taper and optical mirror for confining the optical mode. The OMC is designed to couple the breathing mode $\Gamma$-point to the $X$-point of a TE-like optical mode drawn from the air band. The optical band is pushed up (lowered) with a corresponding increase (decrease) in the proportion of air to silicon within a single unit cell. In the phonon waveguide region (Fig. \ref{FigTaper}), a shrinking $\mathrm{w}_\mathrm{r}$ pushes the optical band upwards and out of the bandgap, creating an photonic crystal mirror. In the phononic mirror region, the shrinking $\mathrm{h}_\mathrm{r}$ has a similar, but gentler effect. This forms a partially-reflective mirror allowing light to couple into the optical cavity. The bands of interest for corresponding unit cells from the different regions of the OMC are shown in Fig. \ref{FigBand}.

The unit cell of 2D phononic shields surrounding the central beam are derived from the previous design \citeSI{mayor2025high}. In our design, the unit cell supports a pseudo-in plane TE bandgap from 177 THz to 226 THz for optics, and a complete phononic bandgap from 6.51GHz to 7.55 GHz. These 2D shields are essential for confining coherent optical photons and mechanical phonons mediating the transduction process, while anchoring the structure to the substrate for better thermalization. 

1D phononic shields are used as supports, allowing aluminum electrodes to extend from contact pads to cover the LN block atop the released silicon membrane. These shields suppress phonon radiation into the substrate from the piezoelectric\citeSI{Chan_2012}. The 1D shield unit cell we designed consists of 80nm of aluminum atop thin-film silicon, and supports a complete phononic bandgap from 6.55 GHz to 7.45 GHz.

\begin{figure*}[h]
\centering
\includegraphics[width=0.25\linewidth]{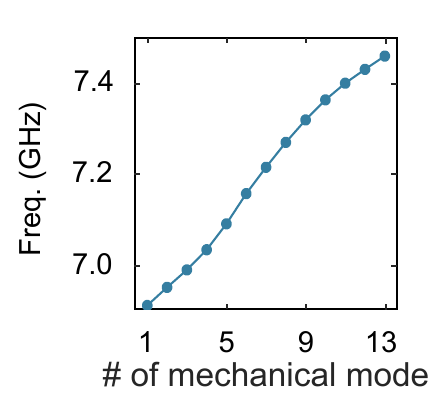}
\caption{\label{Fig_SIFig1e}
{\textbf{Simulated eigenfrequencies for the collective mechanical modes shown in Fig. 1e.} The eigenfrequencies show an almost linear spectra, agreeing with the analogy to a quantum harmonic oscillator.}
}
\end{figure*}

\begin{figure*}[h]
\centering
\includegraphics[width=0.4\linewidth]{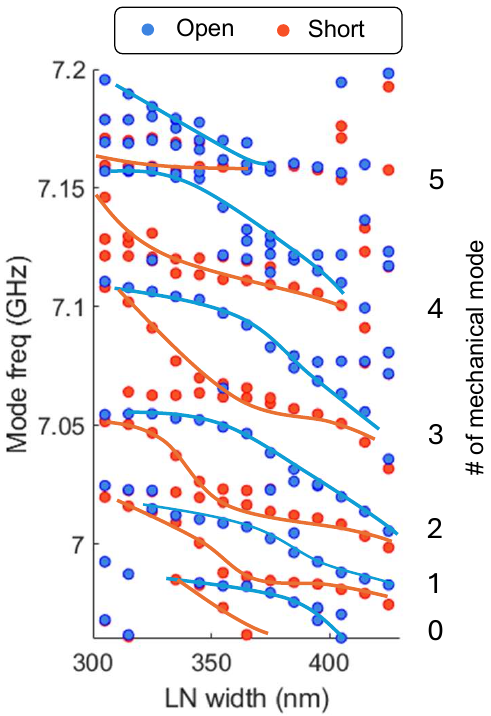}
\caption{\label{SI_LN_w}
\textbf{Simulation of the mechanical modes via open and short boundary condition.}
By changing the terminal settings to voltage (charge), the electric boundary condition is correspondingly shorted (open). The coupling 
strength is therefore extracted based on the frequency change. The mechanical mode labeled from 0 to 5 corresponds to the mode numbering shown in main text Fig. 2c. The lines are plotted for ease of visualization. 
}
\end{figure*}

For the simulation described in Fig. 1e, the mechanical eigenfrequencies follow an almost-linear distribution, as shown in Fig. \ref{Fig_SIFig1e}. To obtain electromechanical coupling/damping rates in simulation, we change the terminal settings to voltage (charge), which changes the electric boundary condition to short (open). The shift of the mechanical eigenfrequencies between boundary conditions is related to the electromechanical coupling of the given mode, and can be described using the LC model for piezoelectric materials\citeSI{Wollack2021}, as shown in Fig. \ref{SI_LN_w}. For the mode 1 shown in main text Fig. 2e, the frequency difference between open and short boundary condition is 20.9 MHz, which corresponds to a $k_\mathrm{eff}^2=\frac{\pi^2}{8}(\frac{\omega_c^2}{\omega_v^2}-1)$=0.74\% \citeSI{Dahmani2020}. With a separate low-frequency admittance simulation, we extract the electromechanical coupling rates assuming an 700 $\Omega$ resonator impedance. The electromechanical damping rate is calculated for a  $Z_0 =$ 50 $\Omega$ probe, using the equation \ref{SI_gem}. Key parameters for the full transducer simulation are summarized in Fig. \ref{SI_POM_summary}.

\begin{table}[H]
    \centering
    \caption{
    \textbf{Summary of the key parameters in the full transducer simulation.}
    $\gamma_\mathrm{em}$ is calculated assuming a 50 $\Omega$ probe and the $g_\mathrm{em}$ is calculated assuming a 700 $\Omega$ resonator impedance. 
    }
    \label{SI_POM_summary}
    \begin{tabular}{|c|c|c|c|c|c|}
        \hline
        Parameters & $Q_\mathrm{op,i}$ & $Q_\mathrm{m}$ & $g_\mathrm{om}$
        & $\gamma_\mathrm{em}$ & $g_\mathrm{em}$\\
        \hline
        Simulations & $7\times10^6$ & $1.4\times10^5$ & $2\pi\times$594 kHz
        & $2\pi\times$18 kHz & $2\pi\times$21 MHz\\
        \hline
    \end{tabular}
\end{table}

\section{Measurement setup}
\subsection{Room temperature experimental setup}
\begin{figure*}[t]
\centering
\includegraphics[width=1\linewidth]{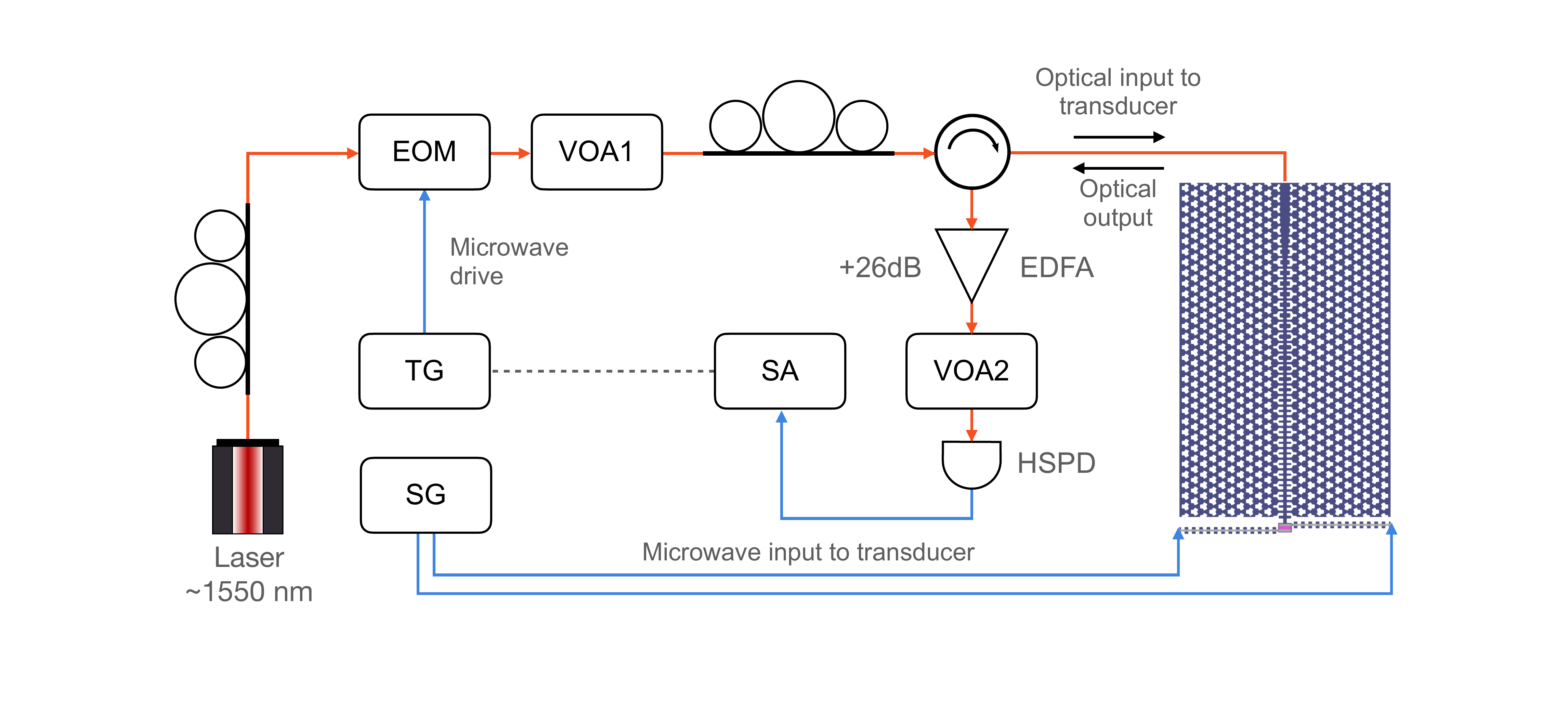}
\caption{\label{FigRTSetup}
\textbf{Optical and microwave setup for characterizing room-temperature electromechanical conversion efficiency.} HSPD: high-speed photodiode. VOA: variable optical attenuator. SA: spectrum analyzer. TG: tracking generator. SG: microwave signal generator. EDFA: erbium-doped fiber amplifier. EOM: electro-optic modulator. Red lines trace optical signal paths while blue lines trace microwave signal paths.
}
\end{figure*}

Our measurement setup for room-temperature transduction is shown in Fig. \ref{FigRTSetup}. We use a telecom laser to supply the optical pump for the transducer (frequency $\omega_\mathrm{L}$). An EOM is driven by a tracking generator (TG) to drive the input optical sideband at $\omega_\mathrm{L} + \omega_\mathrm{m}$, where $\omega_\mathrm{m}$ corresponds to the frequency of the mechanical mode being probed. A microwave tone at $\omega_\mathrm{m}$ is simultaneously applied across the electrodes of the piezoelectric. The output optical signal passes through a circulator and is amplified by a fixed-gain erbium-doped fiber amplifier (EDFA). This signal is attenuated by a variable optical attenuator to avoid saturating or damaging the high-speed photodiode (HSPD). The HSPD signal is read out by a spectrum analyzer (SA) which is paired to the TG. Key parameters for transducer measurements are summarized in Table. \ref{tab:fig4_data}.

\begin{table}[htbp]
    \centering
    \caption{
    \textbf{Table data for Fig.~\ref{Fig4}f and Fig.~\ref{Fig4}g.}
    Summary of the measured key parameters in Fig. 4. The electromechanical coupling rates is calculated using $g_\mathrm{em}^* = \frac{1}{2}\sqrt{\gamma_\mathrm{em}\omega_\mathrm{m}}\sqrt{\frac{Z_\mathrm{c}}{Z_0}}$, assuming a 700 $\Omega$ resonator impedance. 
    }
    \label{tab:fig4_data}
    \begin{tabular}{|c|c|c|c|c|c|}
        \hline
        Mode Number
        & $\gamma_\mathrm{m}/2\pi$
        & $g_{\mathrm{om}}/2\pi$
        & $\gamma_{\mathrm{em}}/2\pi$
        & $g_{\mathrm{em}}^*/2\pi$
        & $\eta_{\mathrm{em}}$ \\
        \hline
        1 & $4.2\,\mathrm{MHz}$ & $540\,\mathrm{kHz}$
          & $1.5\,\mathrm{kHz}$ & $5.9\,\mathrm{MHz}$
          & $3.5 \times 10^{-4}$ \\
        2 & $6.0\,\mathrm{MHz}$ & $446\,\mathrm{kHz}$
          & $4.6\,\mathrm{kHz}$ & $10.5\,\mathrm{MHz}$
          & $6.2 \times 10^{-4}$ \\
        \hline
    \end{tabular}
\end{table}

\subsection{Fridge setup}
We package the transducer chip and the wire-bonded multi-mode microwave resonator by attaching them to a PCB with RF connectors. The PCB sits on top of the copper mount of the mixing chamber plate in the dilution refrigerator. An external coil is used to tune the frequency of the microwave resonator. The amplitude and phase measurement of the microwave S-parameter are shown in Fig. \ref{Fig_SIFig5}, featuring an overcoupled microwave resonator. To calibrate the transduction efficiency, we perform the calibration-free method by measuring all four S parameters \citeSI{Zhao2025}. With $S_\mathrm{oo} = -9.6$dB, $S_\mathrm{oe} = -35.3$dB, $S_\mathrm{eo} = -79.6$dB, $S_\mathrm{ee} = -49.8$dB and optical kappa ratio 19.5\%, we extract the internal transduction efficiency as 0.85\%.

\begin{figure*}
\centering
\includegraphics[width=1\linewidth]{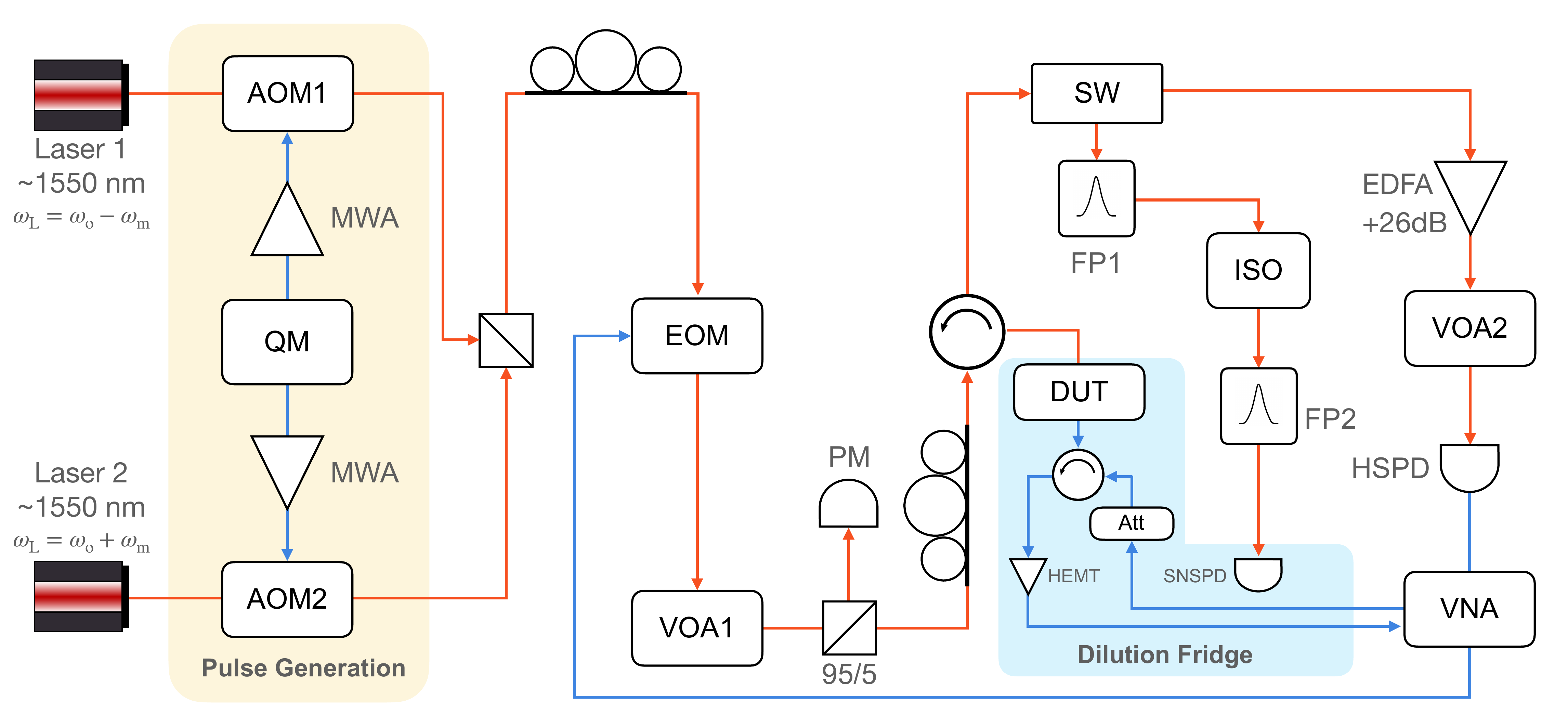}
\caption{\label{FigCryoSetup}
\textbf{Abridged optical and microwave setup for characterizing cryogenic microwave-to-optical transduction.} DUT: device under test, i.e. our 2D transducer. HSPD: high-speed photodiode. SNSPD: superconducting nanowire single photon detector. PM: power meter. VOA: variable optical attenuator. VNA: vector network analyzer. EDFA: erbium-doped fiber amplifier. AOM: acousto-optic modulator. EOM: electro-optic modulator. VNA: vector network analyzer. SW: optical switch. ISO: optical isolator. FP: Fabry-Perot filter. QM: Quantum Machines OPX microwave synthesizer and readout module. MWA: microwave amplifier. HEMT: high-electron-mobility transistor amplifier. Red lines trace optical signal paths while blue lines trace microwave signal paths. Not shown are the laser stabilization components, and in-fridge microwave wiring is summarized.
}
\end{figure*}

\begin{figure*}[h]
\centering
\includegraphics[width=0.7\linewidth]{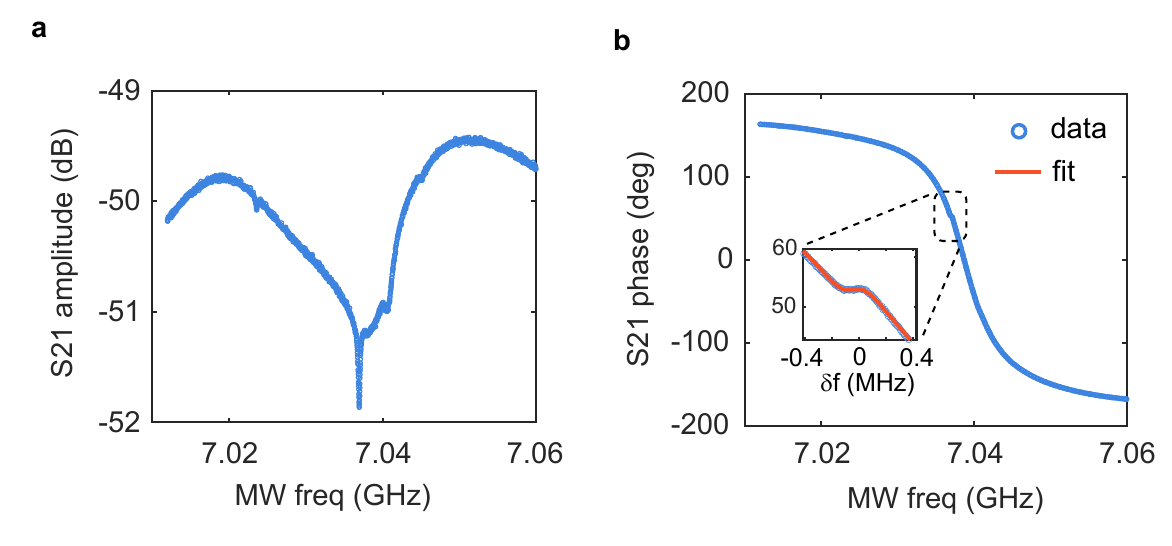}
\caption{\label{Fig_SIFig5}
{\textbf{Amplitude and phase measurement of the microwave S-parameter in Fig. 5b}}
\textbf{a}, Measured amplitude response. The overall background on the microwave spectra makes the amplitude fitting not reliable. 
\textbf{b}, Measured phase response, showing an over-coupled microwave resonator coupled to a mechanical mode. The fit here is with the same fitting parameters used in Fig. 5b, showing an electromechanical coupling rate of $g_{\mathrm{em}}/2\pi = 0.17~\mathrm{MHz}$.
}
\end{figure*}

The same optical setup is used for characterizing the transducer's efficiency and thermometry, shown in Fig. \ref{FigCryoSetup}. CW transduction efficiency measurements utilize one of the two lasers, and the optical switch path leading to the EDFA and HSPD. For thermometry via sideband asymmetry, both lasers are utilized: one is tuned to the blue sideband, while the other is tuned to the red sideband. The optical switch path leading to the cascaded Fabry-Perot filters and SNSPD is used. Heating (blue) and cooling (red) optical pulses are formed by driving AOM1 and AOM2 separately with microwave pulses produced by a Quantum Machines OPX microwave synthesizer.

\section{Device fabrication}

\subsection{OMC only}
We fabricate OMC cavities on a thin film silicon-on-insulator substrate (220 nm device layer, 3~µm buried thermal silicon oxide, 725~µm silicon handle). We pattern the OMC mask using electron-beam lithography (EBL) on CSAR 6200.13 resist, followed by a silicon etch in an inductively-coupled plasma reactive ion etcher. Hydrogen bromide (HBr) and chlorine (Cl$_2$) produce the plasma ion species. The sample is stripped of resist in piranha, and the thin-film silicon structures are released in 49\% hydrofluoric acid (HF), producing a 3~µm undercut.

\subsection{Full transducer}
\begin{figure*}
\centering
\includegraphics[width=1\linewidth]{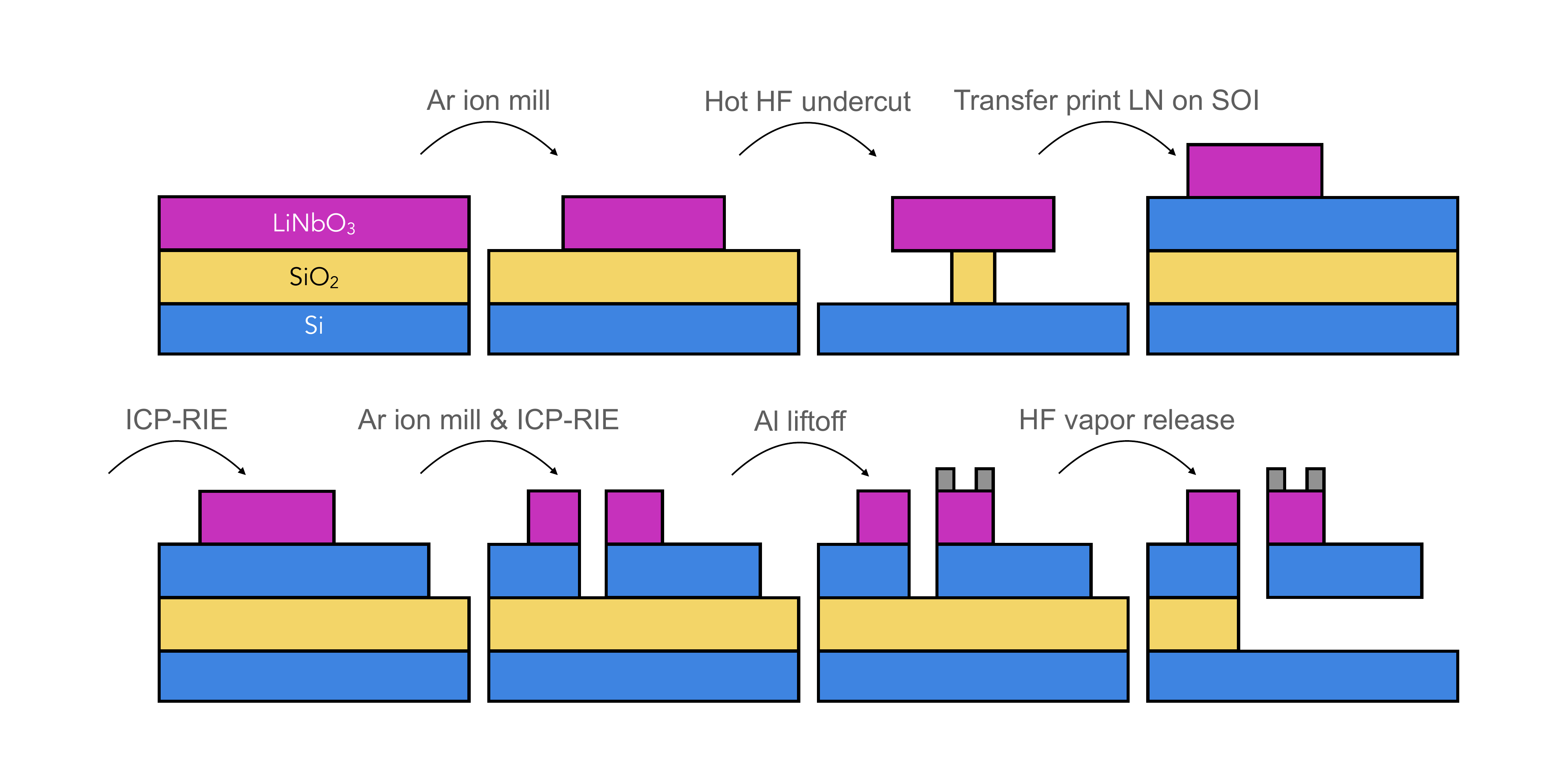}
\caption{\label{FigFab}
{\textbf{Transducer fabrication.} Fabrication process of full transducer chip, including the integration of LN via transfer print.}
}
\end{figure*}

Our transducer's fabrication utilizes the heterogenous integration of LN onto an SOI chip. The fabrication process is summarized in Fig. \ref{FigFab}. We start with an LN-on-insulator chip consisting of 400 nm of top-layer MgO-doped X-cut LN on a 3~µm thick thermal silica layer, atop a 400~µm thick silicon handle. The LN film is thinned to 250 nm with argon ion milling. The LN block and alignment marks are patterned with electron beam lithography (EBL) on hydrogen silsesquioxane (HSQ) resist. The pattern is transferred to LN with a 265 nm etch using argon ion milling. The HSQ is stripped in buffered oxide etchant (BOE, 6:1), and the LN is then cleaned in hot piranha (80 ºC, 3:1 by volume of 96\% sulfuric acid and 30\% hydrogen peroxide).

We then prepare the LN for transfer printing\citeSI{Meitl2005, Jiang2023} onto the SOI chip. The LN is undercut by etching the silica in 5\% HF at 43 ºC for 22 minutes. This produces around 4.5~µm of undercut in the thermal oxide. The released LN pattern is then transferred to a SOI chip via PDMS transfer printing. The chip is annealed for 8 hours at 500 ºC to relieve film stress and improve the LN-Si adhesion. This is followed by a clean in piranha solution. The transfer print causes a $\sim$0.1\% relative displacement between separated patterns. To correct for this, the transferred LN pattern carries alignment marks within 50~µm of every silicon OMC device, to enable sub-30 nm alignment precision on EBL masks after the transfer print. 

Excess LN is removed with an EBL mask patterned in CSAR 6200.18, followed by an argon ion mill. The remaining LN is the block as shown in the main text Fig \ref{Fig2}a. Minimizing the amount of LN present for subsequent EBL masks leads to an increase in pattern fidelity. We noticed this improvement after exchanging the silicon patterning and LN etch steps over several iterations of our devices. A deliberate slight overetch of the LN mask also transfers the alignment marks from the LN into the silicon device layer.

We then pattern the silicon OMCs as described in the previous section. Their alignment is referred to marks created in the silicon during the previous ion milling step. The sample is stripped of resist in piranha, followed by 2\% HF. Subsequently, the aluminum electrodes are defined using a liftoff process. We pattern electrodes in another CSAR 6200.13 mask using EBL, then deposit aluminum in an evaporator at 3 angles (-66º, 0º, +66º), which enables the metal to adhere to the steep sidewalls of the LN. The total aluminum thickness is 80 nm. Excess aluminum is lifted off in an overnight 80ºC N-Methyl-2-pyrrolidone (NMP) soak.

For transducers utilizing aluminum electrodes, we undercut the silicon membrane using anhydrous HF vapor. This method permits selective isotropic etching of the buried oxide layer while minimally etching the aluminum layer. 49\% HF is unsuitable as it tends to strip and etch the aluminum completely from the silicon surface. However, the liquid HF etch tends to produce a cleaner surface than using the vapor etch. This may lead to less optical absorption and reduced surface TLS density, motivating our future endeavors to utilize niobium electrodes that are compatible with the wet etch.

\bibliographystyleSI{naturemag}
\bibliographySI{SIref}

\end{document}